\documentclass[conference,draftcls,onecolumn]{IEEEtran}

\usepackage{amsmath,amsfonts,amssymb,bbm,amsthm} 
\usepackage{multirow}
\usepackage{extarrows}
\usepackage{algorithm}
\usepackage{algorithmic}
\usepackage{enumerate}
\usepackage{threeparttable}
\ifCLASSINFOpdf
\usepackage[pdftex]{graphicx}
\else
\fi
\usepackage{float}
\usepackage{cite}
\usepackage{epsfig}
\usepackage{mathrsfs}
\usepackage{array}
\usepackage{subfig}
\usepackage{verbatim}
\usepackage{setspace}
\usepackage{nicefrac}
\usepackage{arydshln}
\usepackage{changepage}
\usepackage[dvipsnames]{xcolor}
\usepackage{url}

\DeclareGraphicsExtensions{.eps,.pdf,.jpg,.png}

\DeclareMathOperator*{\argmax}{argmax}
\DeclareMathOperator*{\argmin}{argmin}

\newtheorem{theorem}{Theorem}
\newtheorem{lemma}{Lemma}

\newtheorem{definition}{Definition}

\newtheorem{assumption}{Assumption}
\allowdisplaybreaks

\makeatletter
\newenvironment{breakablealgorithm}
{
	\begin{center}
		\refstepcounter{algorithm}
		\hrule height.8pt depth0pt \kern2pt
		\renewcommand{\caption}[2][\relax]{
			{\raggedright\textbf{\ALG@name~\thealgorithm} ##2\par}%
			\ifx\relax##1\relax 
			\addcontentsline{loa}{algorithm}{\protect\numberline{\thealgorithm}##2}%
			\else 
			\addcontentsline{loa}{algorithm}{\protect\numberline{\thealgorithm}##1}%
			\fi
			\kern2pt\hrule\kern2pt
		}
	}{
		\kern2pt\hrule\relax
	\end{center}
}
\makeatother

\begin{document}

\title{Towards Fast-Convergence, Low-Delay and
	Low-Complexity Network Optimization}

\author{Sinong Wang and Ness Shroff
\thanks{Sinong Wang  and Ness Shroff were with the Department
of Electrical and Computer Engineering, The Ohio State University, Columbus,
OH, 43210 USA e-mail: \{wang.7691, shroff.11\}@osu.edu}}

\maketitle

\begin{abstract}
Distributed network optimization has been studied for well over a decade. However, we still do not have a good idea of how to design schemes that can simultaneously provide good performance across the dimensions of utility optimality, convergence speed, and delay. To address these challenges, in this paper, we propose a new algorithmic framework with all these metrics approaching optimality. The salient features of our new algorithm are three-fold: (i) fast convergence: it converges with only $O(\log(1/\epsilon))$ iterations that is the fastest speed among all the existing algorithms; (ii) low delay: it guarantees optimal utility with finite queue length; (iii) simple implementation: the control variables of this algorithm are based on virtual queues that do not require maintaining per-flow information. The new technique builds on a kind of inexact Uzawa method in the Alternating Directional Method of Multiplier, and provides a new theoretical path to prove global and linear convergence rate of such a method without requiring the full rank assumption of the constraint matrix.
\end{abstract}


\IEEEpeerreviewmaketitle

\section{Introduction}

Consider a fixed data network shared by $F$ end-to-end flows. Each flow $f$ is described by its source-destination node pair and associated utility function, without a priori established routes. The nodes within the network cooperate by forwarding each others' packets toward their destinations. The network optimization problem is \emph{how does one jointly choose the end-to-end data rate $x_f$ of each flow $f$, the schedule for each link and the link rate for each flow to maximize the network utilities defined as}
\begin{equation}\label{eq:crossopt}
\max\sum\limits_{f=1}^F U_f(x_f) \text{ s.t. } [x_f]\in\Lambda,
\end{equation}
where $\Lambda$ is the capacity region of data network, dependent on the limited power resources and interference among concurrent transmissions. The optimization problems of the above form plays a key role in resource control and optimization for both wireline and wireless networks. 

In distributed network optimization, each iteration of the algorithm corresponds to one communication among different nodes, which could require a very large amount of information exchange overhead. Therefore, one important metric to measure the performance of algorithm is the convergence speed, i.e., how many iterations are required to obtain an $\epsilon-$accurate solution. In addition, other important metrics are utility and the physical queue length in steady state, which measures the throughput and transmission delay that is achieved by the algorithm. 

\subsection{Existing Algorithms}

The large body of work (see, e.g.,~\cite{tassiulas1992stability,neely2003power,lin2004joint,stolyar2005maximizing,eryilmaz2006joint,lin2006utility,liu2016heavy,liu2016achieving,wei2013distributed,liu2013distributed,liu2016joint}, and~\cite{lin2006tutorial} for a survey) in this area has given rise to several efficient and distributed control algorithmic frameworks. We first review the state-of-the-art of all the existing algorithms.

\textbf{First-order dual decomposition method:} This kind of algorithm applies the subgradient descent method to the dual function of problem (\ref{eq:crossopt}) and lead to a beautiful queue-length-based control algorithmic (QCA) framework, based on which the components of congestion control, routing and scheduling are naturally coupled by queuing states~\cite{lin2004joint,eryilmaz2006joint}. However, the classical QCA method achieves an $O(1/K)$ utility optimality gap at an expense of $O(K)$ steady-state queue-length, where $K>0$ is a system parameter. Hence, a small utility gap will yield a large queuing delay. Significant efforts have been made to improve this tradeoff including the development of virtual queue techniques~\cite{kunniyur2001analysis,athanasopoulou2013back}, the threshold-based packeting-dropping scheme~\cite{huang2011delay} and the $[O(1/K),O(\sqrt{K})]$ tradeoff produced by recent momentum-based methods~\cite{liu2016heavy,liu2016achieving}. Due to the nonsmoothness of dual function and the subgradient nature, all the above methods suffer from a slow convergence that requires $O(1/\epsilon^2)$ iterations to obtain an $\epsilon-$accurate solution. 

\textbf{Second-order Newton method:} To improve the convergence speed, there have been many attempts in obtaining new algorithms by applying the second-order method~\cite{wei2013distributed,liu2013distributed,liu2016joint}. Compared with the first-order method, this kind of algorithm has a faster convergence rate, i.e., $O(\log^2(1/\epsilon))$ iterations (three-level convergence structure with interior point, Newton and matrix splitting method). However, it has several limitations: (i) the complexity of computing the Hessian inverse in the second-order method is quite high and does not scale well with the network size; (ii) a worse utility-delay tradeoff $[O(1/K),$ $O(K^2)]$ in~\cite{liu2016joint}; (iii) it cannot efficiently handle the wireless interference channel. For example, in the algorithm~\cite{liu2013distributed}, even the number of variables (time sharing parameters) in the control plane is exponentially large.

\textbf{Proximal method:} The proximal method was first introduced in the work~\cite{lin2006utility} to tackle the oscillation problem in a network optimization problem with given routing paths. Unlike the QCA method, it adds a quadratic regularizer in the routing component to stabilize the solution, and is proven to be the first algorithm to break the existing utility-delay tradeoff that offers both the zero utility optimality gap and finite queue length. Recently, the work~\cite{yu2017new} generalizes this idea to the scenario of dynamic routing and designs a new backpressure routing algorithm for wireline network. They prove that the proximal method not only exhibits the feature of low-latency, it also offers an improved convergence speed of $O^*(1/\epsilon)$\footnote{Here the $O^*(1/\epsilon)$ means that the convergence rate is in the ergodic sense. A sequence $\{a_n\}$ converges with ergodic rate $O^*(1/\epsilon)$ if $\frac{1}{N}\sum_{n=1}^Na_n=O(1/N)$, with rate $O(1/\epsilon)$ if $a_n=O(1/N)$.}.

It can be observed that all the existing algorithms sacrifice the performance of one or more metrics to improve the others. In particular, the  slow convergence of all these algorithms will result in large information exchange overhead. The key question that we aim to answer in this paper is that: \emph{is it possible to develop a joint congestion control, routing and scheduling algorithm with the fast convergence speed, routing complexity as low as the first-order method and delay as low as the proximal method?}

\begin{table*}[!t]
	\centering
	\vspace{-0.1in}
	\caption{Comparison of Existing Algorithms in Network Optimization}\label{tab:result}
	\footnotesize
	\begin{threeparttable}
		\begin{tabular}{|c|c|c|c|c|c|}
			\hline
			&Optimality gap & Queue-length &  Convergence speed & Routing complexity & Scheduling complexity\tnote{1}\\
			\hline
			Dual decomposition method & $O(1/K)$ & $O(K)$ & $O(1/\epsilon^2)$ & $O(F)$ & poly$(L,F)$\\
			\hline
			Proximal method & optimal& $O(1)$ & $O^*(1/\epsilon)$ & $O(F\log(F))$ & unknown \\
			\hline
			Second-order  method & $O(1/K)$ & $O(K^2)$ & $O(\log^2(1/\epsilon))$ & $O(F^2+L^2)$ & exp$(L,F)$ \\
			\hline
			Momentum method& $O(1/K)$ & $O(\sqrt{K})$ & $O(1/\epsilon^2)$ & $O(F)$ & poly$(L,F)$ \\
			\hline
			\textbf{Our new method} &  optimal& $O(1)$& $O(\log(1/\epsilon))$ & $O(F\log(F))$ & poly$(L,F)$ \\
			\hline
		\end{tabular}
		\begin{tablenotes}
			\scriptsize
			\item[1] The scheduling complexity derives from the traditional node-exclusive interference model.
			\item[2] Momentum method refers to heavy-ball method and Nesterov's accelerated method.
		\end{tablenotes}
	\end{threeparttable}
	\vspace{-0.2in}
\end{table*}

\subsection{Our Results}

In this paper, we positively answer this open question and propose a new algorithmic framework. The comparison of our algorithm and the existing schemes in a $L-$links and $F-$flows network are listed in TABLE~\ref{tab:result}. One can see that our algorithm offers the \textbf{fastest} convergence speed, \textbf{optimal} utility,  \textbf{finite} queue length, and \textbf{low} routing and scheduling complexity compared with all the existing methods. The rationale behind our algorithm design is to utilize the Alternating Directional Method of Multiplier (ADMM), first appeared in~\cite{gabay1976dual}. Our key idea is to reformulate the joint scheduling-routing-congestion control problem as a $2-$block separable optimization problem, and apply the ADMM to the Augmented Lagrangian function of problem (\ref{eq:crossopt}), which then allows us to obtain an optimization framework with a layered structure and only a limited degree of cross-layer coupling. 

However, due to a number of technical challenges, developing an ADMM-based method is highly non-trivial. First, the ADMM's focus is on minimizing the Augmented Lagrangian function that is the summation of original utility function and a quadratic penalty function of the constraints. It will produce a routing-scheduling problem with a non-separable objective function regarding the rate vector among different links. Therefore, it is difficult to be solved in a low-complexity and distributed manner. Second, the structure of this method is substantially different from both the dual decomposition method and the proximal method. For example, the form of congestion control, routing component, and the coupling among the different layers are different. Hence, the analytical techniques used in existing methods for utility optimality and queue stability are not applicable. Third, in a wireless network with interference constraints, unlike the clear relationship between the linear program-based scheduling problem in the dual decomposition method and the combinatorial optimization problem, i.e., maximum weighted matching~\cite{lin2004joint,sharma2006complexity}, it is unclear how to solve the new scheduling problem derived from the ADMM-based decomposition.

The main contribution of this paper is that we develop a new algorithmic framework that addresses the aforementioned challenges. The detailed results and technical contributions of this paper are as follows:
\begin{itemize}
	\item We utilize a kind of inexact Uzawa method of Alternating Directional Method of Multiplier~\cite{bramble1997analysis,zhang2011unified} to approximately solve a local second-order approximation of the Augmented Lagrangian function with respect to the link rates. This technique will yield a routing and scheduling problem with a separable quadratic objective function and a constraint set defined by a convex hull of feasible link rate vectors.
	\item We establish the utility optimality and finite queue length of our proposed framework. In particular, we show that, as the algorithm keeps running, the network utility gap will vanish, while the queue lengths in each node are bounded throughout by a finite constant. This result is much stronger than the best tradeoff $[O(1/K),O(\sqrt{K})]$ of the traditional QCA framework. Moreover, we prove that our new algorithmic framework converges at a global and linear rate that obtains an $\epsilon-$accurate solution with only $O(\log(1/\epsilon))$ number of iterations, which is faster than the existing second-order methods.
	\item We provide several algorithms to implement the new routing and scheduling problem in our proposed framework. More precisely, for the wireline network, we show that the new routing problem can be solved in a distributed manner and in $O(F\log(F))$ time within each link, which is much lower than $O(F^2+L^2)$ complexity of the second-order method in~\cite{liu2016joint}. For the wireless networks with interference constraints, we show that the complexity of solving our new scheduling problem is equivalent to the classical MaxWeight scheduling. This result not only implies a deep connection between these two problems, but also paves a path to use the existing algorithms~\cite{sharma2006complexity,jiang2010distributed} of MaxWeight scheduling to solve this new problem.
\end{itemize}

One technical contribution independent of interest is the global and linear convergence rate of our proposed algorithm. As mentioned earlier, this  algorithm is indeed applying an inexact Uzawa method of ADMM to the optimization problem of the form $\min f(\mathbf{x})+g(\mathbf{y}),\text{ s.t. }\mathbf{Ax}+\mathbf{By}=\mathbf{b}$. All the existing global and linear convergence results~\cite{lin2015global,deng2016global,davis2017faster} of this generalized ADMM requires an assumption that one of the constraint matrices is of full rank. However, in our problem, both matrices $\mathbf{A}$ and $\mathbf{B}$ do not satisfy this condition. We provide a new technical path to overcome this challenge. The critical technical step is to estimate the distance from the primal and dual iterates of ADMM to the optimal solution set by the distance to an inscribed polyhedron of the optimal set. This enables us to utilize the isolated calmness of polyhedral mapping to upper bound such distance by certain amount of constraint violation.

The remainder of this paper is organized as follows. In
Section 2, we introduce the network model and problem formulation. Section 3 presents our proposed algorithmic framework and the main results. In Section 4, we provide the detailed theoretical analysis of convergence speed and queuing stability. Section 5 develops the algorithms for the principal components of our framework.
Section 6 presents numerical results. Section 7 provides some discussions and Section 8 concludes this paper. Due to the space limit, all the proofs are listed in Appendix.

\section{Problem Statement}

\subsection{Network Model}

\begin{figure}[!t]
	\centerline{ \includegraphics[angle=0,width=0.8\textwidth]{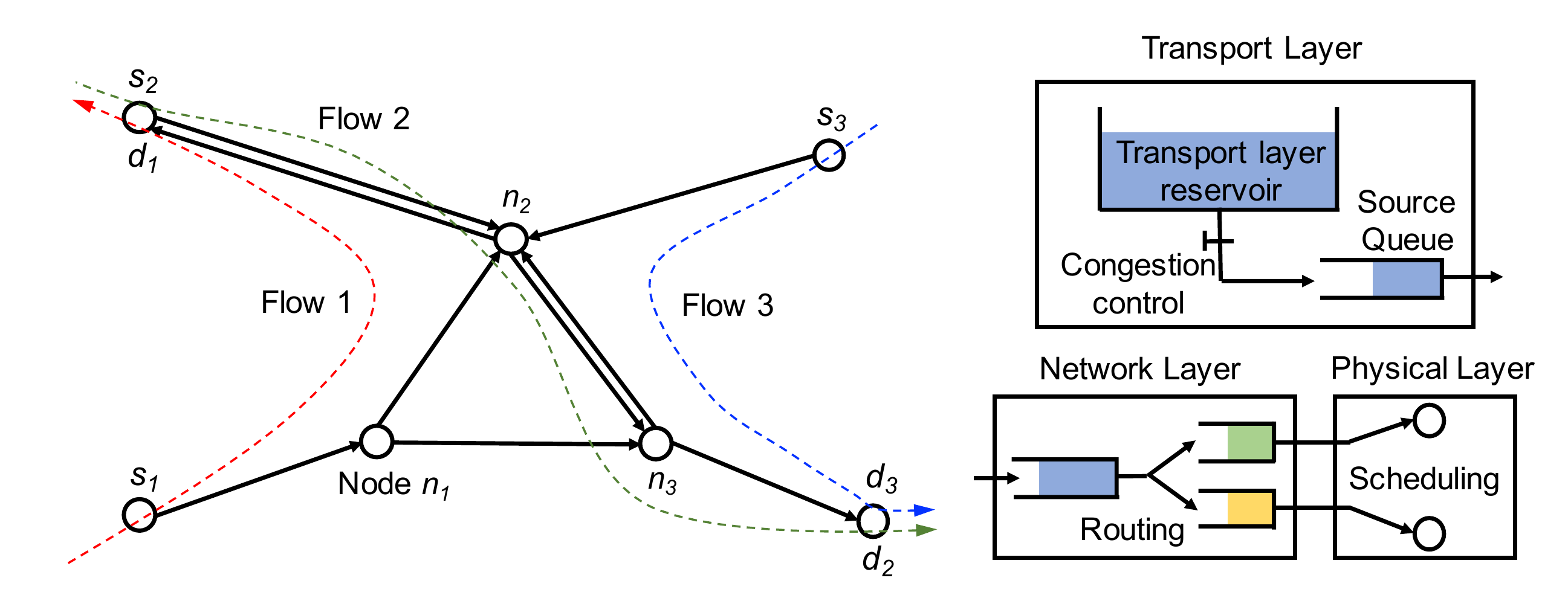}}
	\caption{Illustrative example of model.}
	\label{fig:model}
\end{figure}

We consider a slotted communication network system with time slot units being indexed by $t=1,2,\ldots$. As shown in Fig.~\ref{fig:model}, we represent the network by a \emph{directed} graph $\mathcal{G}=\{\mathcal{N},\mathcal{L}\}$, where $\mathcal{N}$ is the set of nodes and $\mathcal{L}$ is the set of edges. Let $|\mathcal{N}|=N$ and $|\mathcal{L}|=L$.  For each node $n$, denote the sets of its incoming links and outgoing links as $\mathcal{I}(n)$ and $\mathcal{O}(n)$, respectively. Let deg$(n)$ be the number of adjacent links of node $n$. We define $\text{Tx}(l)$ and $\text{Rx}(l)$ as the transmitting and receiving node for each edge $l$. There are $F$ end-to-end sessions in the network, indexed  by $f\in\mathcal{F}\triangleq\{1,2,\ldots,F\}$. Each session $f$ has a source node $s_f$ and a destination node $d_f$ in the node set $\mathcal{N}$. To avoid triviality, suppose that different sources are located at different nodes.

\subsection{Congestion Control}
Let scalar $x_f$ be the injection rate of session $f$ with which data is sent from $s_f$ to $d_f$, possibly via multiple hops and multiple paths. We assume that injection rate $x_f$ is bounded in $[m_f,M_f]$. Associated with each flow $f$ is a utility function $U_f(x_f)$, which reflects the ``utility'' to session $f$ when it can transmit at rate $x_f$. We assume that the utility function $U_f(\cdot)$ satisfy the following conditions.
\begin{assumption} \emph{(Utility function)}\label{ass:obj}
	For each session $f$, the utility function $U_f(\cdot)$ is a nondecreasing and concave function in the interval $[m_f,M_f]$.
\end{assumption}
The use of such utility functions is common in the congestion control literature to model fairness. For example, these conditions hold for the following two typically used utility functions: (i) weighted proportionally fair utilities $U_f(x_f)=w_f\log(x_f)$, where $w_f,f=1,\ldots,F$ are the weights; (ii) general weighted proportionally fair utilities,
\begin{equation}
U_f(x_f)=w_f\frac{x^{1-\gamma}}{1-\gamma}, \gamma>0.
\end{equation}
Note that these two examples are also strictly concave functions.

\subsection{Routing and Scheduling}

For each edge $l$ in the set $\mathcal{L}$, suppose that $l=(m,n)$ and the data is transmitted from node $m$ to node $n$. Let $r_{l}^d$ represent the amount of capacity on link $l$ that is allocated for data towards destination $d$. In the sequel, we call it the link rate for simplicity. The set of destination nodes are defined as $\mathcal{D}=\{d_f,f\in\mathcal{F}\}$, and let $|\mathcal{D}|=D$.  Then we can describe the capacity region of the data network.
\begin{definition} \emph{(Capacity Region~\cite{tassiulas1992stability,neely2003power})}
	The capacity region $\Lambda$ of the network is the largest set of injection rate vector $[x_f]_{f\in\mathcal{F}}$ for which there exists a link rate vector $[r_{l}^d]^{d\in\mathcal{D}}_{l\in\mathcal{L}}$ that satisfies the following constraints.
	\begin{enumerate}
		\item Flow conservation: for each destination $d$ in $\mathcal{D}$, each node $n$ in $\mathcal{N}\backslash\{d\}$,
		\begin{equation}\label{eq:flowconserve}
		\sum\limits_{f\in\mathcal{F}}x_f\mathbbm{1}_{\{s_f=n,d_f=d\}}+\sum\limits_{l\in\mathcal{I}(n)}r_l^d=\sum\limits_{l\in\mathcal{O}(n)}r_l^d,
		\end{equation}
		where $\mathbbm{1}_{\{\cdot\}}$ is an indicator function that takes the value $1$ if $s_f=n, d_f=d$ and $0$ otherwise.
		\item Capacity constraint: for each link $l\in\mathcal{L}$ and $d\in\mathcal{D}$,
		\begin{equation}\label{eq:capacitycons}
		\left[\sum\limits_{d\in\mathcal{D}}r_l^d\right]\in\mathcal{C}\triangleq \emph{Conv}(\Gamma), r_l^d\geq 0, 
		\end{equation}
		where $\Gamma=\{\mathbf{r}^{(1)},\mathbf{r}^{(2)},\ldots,\mathbf{r}^{(I)}\}$
		is the set of feasible link rate vectors, and \emph{Conv}$(\cdot)$ represents the convex hull operation.
	\end{enumerate}
\end{definition}


\subsection{Queue Stability}

We use $Q_n^{d}[t]$ to denote the length of the physical queue that are destined for node $d$, waiting
for service at node $n$ in time slot $t$. For each $d\in\mathcal{D}$ and $n\in\mathcal{N}\backslash d$, the evolution of physical queue length is given by
\begin{align}
Q_n^{d}[t]=&\left[Q_n^{d}[t-1]-\sum\limits_{l\in\mathcal{O}(n)}r_l^d[t]\right]_+ +\sum\limits_{l\in\mathcal{I}(n)}\hat{r}_l^d[t]+\sum\limits_{f\in\mathcal{F}}x_f[t]\mathbbm{1}_{\{s_f=n,d_f=d\}},\label{eq:phyQupdate}
\end{align}
where $[\cdot]_+\triangleq\max\{\cdot,0\}$. The rate $r_l^d[t]$ is the capacity provided to $d$-destined packets over link $l$ in time slot  $t$ and the rate $\hat{r}_l^d[t]$ is the actual used capacity over link $l$ for $d$-destined packets in time slot $t$. We have $\hat{r}_l^d[t]\leq r_l^d[t]$ since node $n$ may have less than $ r_l^d[t]$ amount of data to transmit for destination $d$. Note that the definition of $Q_n^{d}[t]$ is only used to measure the delay performance of our algorithm. The actual operation of our algorithm does not require this information (details in Section~\ref{sec:frame}).
\begin{definition} \emph{(Network Stability)}
	Under a congestion control, routing and scheduling scheme, we say that the network is stable if the sum of queue lengths in steady state remains finite.
	\begin{equation}
	\lim\sup_{t\rightarrow\infty}\sum\limits_{d\in\mathcal{D}}\sum\limits_{n\in\mathcal{N}\backslash d}Q_n^d[t]<+\infty.
	\end{equation}
\end{definition}

\subsection{Problem Formulation}

Our objective is to develop a joint congestion
control, routing and scheduling algorithm to maximize the total utility $\sum_{f\in\mathcal{F}}U_f(x_f)$, subject to the network capacity constraints. Putting together the models presented earlier leads to the following general multi-commodity network flow formulation.
\begin{align}
&\textbf{JCRS:}\notag\\
&\max\limits_{x_f,r_l^d}\quad \sum_{f\in\mathcal{F}}U_f(x_f)\label{eq:model}\\
& \begin{array}{r@{\quad}l@{}l@{\quad}l}
s.t. &\sum\limits_{f\in\mathcal{F}}x_f\mathbbm{1}_{\{s_f=n,d_f=d\}}+\sum\limits_{l\in\mathcal{I}(n)}r_l^d=\sum\limits_{l\in\mathcal{O}(n)}r_l^d, \forall d,n\in\mathcal{N}\backslash d,\notag\\
&\left[\sum\limits_{d\in\mathcal{D}}r_l^d\right]\in\mathcal{C}, r_l^d\geq 0, \forall d\in\mathcal{D}, l\in \mathcal{L},\notag\\
&m_f\leq x_f\leq M_f, \forall f\in\mathcal{F}.
\end{array} 
\end{align}
Problem (\ref{eq:model}) is a convex program with affine constraints. We make the following standard assumption that is used in all the existing works.
\begin{assumption}\label{ass:slater}
	\emph{(Existence of optimal solutions)} There exists an optimal injection rate vector $[x^*_f]_{f\in\mathcal{F}}$, link rate vector  $[{r^d_l}^*]_{l\in\mathcal{L}}^{d\in\mathcal{D}}$ and the Lagrangian multiplier vector $[{\lambda_n^d}^*]_{n\in\mathcal{N}\backslash d}^{d\in\mathcal{D}}$ in the problem (\ref{eq:model}).
\end{assumption}
Note that the existence of optimal primal and dual solutions can be be guaranteed if a certain constraint qualification such as the Slater condition holds~\cite{rockafellar2015convex}. In what follows, we will investigate a new distributed joint congestion control, routing and scheduling algorithm.

\section{Joint Congestion Control, Routing and Scheduling Framework}

In Section~\ref{sec:frame}, we first introduce our new algorithmic framework. Then, in Section~\ref{sec:frame:res}, we present the main results on the utility optimality, queue stability and the convergence speed of the proposed algorithm.

\subsection{Algorithmic Framework\label{sec:frame}}

The main procedure of our new joint congestion control, routing and scheduling method is described in Algorithm~\ref{alg:frame}.

\begin{breakablealgorithm}
	\caption{New Joint Congestion Control, Routing and Scheduling Framework}\label{alg:frame}
	\renewcommand{\algorithmicrequire}{\textbf{Initialization:}}  
	\renewcommand{\algorithmicensure}{\textbf{Iteration:}} 
	\begin{algorithmic}[1] 
		\REQUIRE ~~\\
		Choose parameters $\rho>0$, $\tau\in[1,\frac{\sqrt{5}+1}{2})$ and $\beta_{m,n}>\text{deg}(m)+\text{deg}(n),\forall (m,n)\in\mathcal{L}$. Set $t=0$. Let both physical and virtual queues be empty at the initial state $Q_n^d[0]=\lambda_n^d[0]=\lambda_n^d[-1]=0, \forall d\in\mathcal{D}$ and $n\in\mathcal{N}\backslash\{d\}$. Let injection rates $x_f[0]=0, \forall f\in\mathcal{F}$ and service rates $r^d_{l}[0]=0,\forall d\in\mathcal{D}, l\in\mathcal{L}$.
		\ENSURE In each time slot $t\geq 1$, repeat the following three  steps.~~\\ 
		\STATE \textbf{Routing and Scheduling:} For each destination $d\in\mathcal{D}$ and node $n\in\mathcal{N}\backslash\{d\}$, calculate the new weight $z_n^d[t]=(1+1/\tau)\lambda_n^d[t-1]-\lambda_n^d[t-2]/\tau$. Let $z_d^d[t]=0,\forall d\in\mathcal{D}$. Then choose the link rate $[r_{l}^d[t],l\in\mathcal{L},d\in\mathcal{D}]$ as the solution to the following quadratic program.
		\begin{align}
		&\max\limits_{r_{m,n}^d}\sum\limits_{(m,n)\in\mathcal{L}} \sum\limits_{d\in\mathcal{D}}(z_m^d[t]-z_n^d[t])r_{m,n}^d-\frac{\rho\beta_{m,n}}{2}(r_{m,n}^d-r_{m,n}^d[t-1])^2\notag\\
		& \begin{array}{r@{\quad}l@{}l@{\quad}l}
		s.t. &\left[\sum_{d}r_{m,n}^d\right]\in\mathcal{C}, r_{m,n}^d\geq 0, \forall (m,n)\in\mathcal{L},d\in\mathcal{D}.\label{eq:nbpscheduling}
		\end{array} 
		\end{align}
		
		\STATE \textbf{Congestion Control:} For each node $s_f$, calculate the injection rate $x_f[t]$ as the solution to the following optimization problem.
		\begin{equation}\label{eq:congestioncontrol}
		\max\limits_{x_f \in[m_f,M_f]} U_f(x_f)-(z_{s_f}^{d_f}[t]+\rho\Delta r_{f}[t])x_f-\frac{\rho}{2}(x_f-x_f[t-1])^2.
		\end{equation}
		where the quantity $\Delta r_{f}[t]$ is given by
		\begin{align}
		\Delta r_{f}[t]=&\sum\limits_{l\in\mathcal{I}(s_f)} \left(r_l^{d_f}[t]-r_l^{d_f}[t-1]\right)-\sum\limits_{l\in\mathcal{O}(s_f)} \left(r_l^{d_f}[t]-r_l^{d_f}[t-1]\right).\label{eq:deltar}
		\end{align}
		
		\STATE \textbf{Virtual Queue Update:} For each destination $d\in\mathcal{D}$ and node $n\in\mathcal{N}\backslash\{d\}$, update the virtual queue length by
		\begin{align}
		\lambda_n^{d}[t]=&\lambda_n^{d}[t-1]-\rho\tau\sum\limits_{l\in\mathcal{O}(n)}r_l^d[t] +\rho\tau\sum\limits_{l\in\mathcal{I}(n)}r_l^d[t]+ \rho\tau\sum\limits_{f\in\mathcal{F}}x_f[t]\mathbbm{1}_{\{s_f=n,d_f=d\}}.\label{eq:virtualQ}
		\end{align}
	\end{algorithmic}
\end{breakablealgorithm}

Some important remarks on Algorithm 1 are in order:

\textbf{Relation to QCA:} In the QCA method~\cite{lin2004joint,eryilmaz2006joint}, the congestion control component has the form of
\begin{equation}
\max\limits_{x_f \in[m_f,M_f]} U_f(x_f)-Q_{s_f}^{d_f}[t]x_f,
\end{equation}
and the routing and scheduling component is given by
\begin{align}
&\max\limits_{r_{m,n}^d}\sum\limits_{(m,n)\in\mathcal{L}} \sum\limits_{d\in\mathcal{D}}(Q_m^d[t]-Q_n^d[t])r_{m,n}^d\notag\\
& \begin{array}{r@{\quad}l@{}l@{\quad}l}
s.t. &\left[\sum_{d}r_{m,n}^d\right]\in\mathcal{C}, r_{m,n}^d\geq 0, \forall (m,n)\in\mathcal{L},d\in\mathcal{D}.\label{eq:bpscheduling}
\end{array} 
\end{align}
Each component in this method is ``loosely'' connected by the physical queue length $Q_n^d[t]$. Similarly, our new algorithm also exhibits a layered structure, however, each component is ``densely'' connected by several quantities including the virtual queue length $\lambda_n^d[t]$, the injection rate $x_f[t]$ and the link rate $r_l^d[t]$. For example, the congestion control in the source node is dependent on both the virtual queue length and the change of link rate $\Delta r_{f}[t]$ in the adjacent links.

\textbf{Quadratic congestion control and routing:} Unlike the QCA method, Algorithm~\ref{alg:frame} contains a separable quadratic function in each component. In~\cite{lin2006utility}, it has been observed that such a $l_2$-regularization in the routing component can resolve the oscillation problem that occurs in traditional backpressure routing (\ref{eq:bpscheduling}). Technically, we will see later that this technique also leads to significant delay reduction and convergence speed up, moreover, it can be derived from a kind of inexact Uzawa method in Alternating Directional Method of Multiplier~\cite{bramble1997analysis,zhang2011unified}. 

\textbf{Virtual queue-based control:} Existing methods such as the dual decomposition and the momentum-based methods require each node to maintain a separate physical queue for each flow, which is usually difficult to implement, especially in large networks. However, one can see that all the operations of congestion control, routing and scheduling in Algorithm~\ref{alg:frame} are based on the virtual queue length $\lambda_n^d[t]$. In practice, each node will maintain a separate virtual queue (i.e., a counter) for each flow going through it and a FIFO queue for storing packets of all the flows going through the corresponding link. This technique can significantly decrease the complexity of the queuing data structures at each node. Detailed implementation can be seen in~\cite{bui2009novel}.

\subsection{Main Results\label{sec:frame:res}}

For notational convenience, we use vectors $\mathbf{x}[t], \mathbf{r}[t],\boldsymbol{\lambda}[t]$ to group all the injection rates, link rates and virtual queue lengths in time slot $t$, respectively. The first result in this paper is on the utility optimality and queue stability of Algorithm~\ref{alg:frame}.
\begin{theorem}\label{thm:optstab} \emph{(Utility optimality and queue stability)} Under the Assumptions~\ref{ass:obj} and~\ref{ass:slater}, the network utility and physical queue length produced by Algorithm~\ref{alg:frame} satisfies
	\begin{align}
	&\lim\sup\limits_{t\rightarrow\infty} \left|\sum\limits_{f\in\mathcal{F}}U_f(x_f[t])-\sum\limits_{f\in\mathcal{F}}U_f(x_f^*)\right|=0,\label{eq:vanobj}\\
	&\lim\sup_{t\rightarrow\infty}\sum\limits_{d\in\mathcal{D}}\sum\limits_{n\in\mathcal{N}\backslash d}Q_n^d[t]<+\infty,\label{eq:finiteQ}
	\end{align}
	where $[x_f^*, f\in\mathcal{F}]$ is the optimal injection rate vector.
\end{theorem}
Theorem~\ref{thm:optstab} says that our proposed algorithm achieves optimal utility while guaranteeing that the physical queue length at each node is a finite constant. This result 
improves the utility-delay tradeoffs of prior works including $[O(1/K),$ $O(K^2)]$ in~\cite{liu2016joint}, $[O(1/K),O(K)]$ in~\cite{lin2004joint} and $[O(1/K),O(\sqrt{K})]$ in~\cite{liu2016heavy,liu2016achieving}. All these methods will produce an unbounded queue length to obtain a vanishing utility optimality gap. 

\begin{theorem}\label{thm:linconv} \emph{(Global and linear convergence rate)} Under Assumptions~\ref{ass:obj} and~\ref{ass:slater} and the assumption that utility function is strictly concave, the Algorithm~\ref{alg:frame} converges at a global and linear rate. More specifically, there exists one of the optimal injection rate vector $\mathbf{x}^*$, link rate vector $\mathbf{r}^*$ and dual variable $\boldsymbol{\lambda}^*$ of the problem (\ref{eq:model})  such that $\|\mathbf{x}[t]-\mathbf{x}^*\|\leq O(c^t)$, $\|\mathbf{r}[t]-\mathbf{r}^*\|\leq O(c^t)$, $\| \boldsymbol{\lambda}[t] - \boldsymbol{\lambda}^* \|\leq O(c^t)$ for all $t\geq 1$, where $c$ is a constant satisfying $0<c<1$.
\end{theorem}
As can be seen in Theorem~\ref{thm:linconv}, to obtain an $\epsilon-$accurate solution, our new algorithm only requires $O(\log(1/\epsilon))$ iterations, or equivalently, solving number of  $O(\log(1/\epsilon))$ congestion control and routing components. This iteration complexity is much less than the traditional first-order method including dual decomposition method with $O(1/\epsilon^2)$ or the proximal method with $O(1/\epsilon)$. Moreover, it is even faster than the three-layered second-order Newton method~\cite{liu2013distributed}.

Currently,  several natural questions arise are: (i) how to design this new joint scheduling-routing-congestion control algorithm? (ii) how to prove the linear convergence rate, optimal utility and finite queue length of this new algorithm? (iii) how to efficiently solve the quadratic congestion control, routing and scheduling component in our new algorithm? In the sequel, we focus on answering these questions.

\section{Theoretical Analysis}

In this section, we first provide some necessary notations and basics in the variational analysis. Then, we will show how to apply the inexact Uzawa method in the Alternating Directional Method of Multiplier to obtain Algorithm~\ref{alg:frame}. Finally, we will prove the technical results stated in Theorems~\ref{thm:optstab} and~\ref{thm:linconv}.

\subsection{Notations and Preliminaries}

We use the bold letter $\mathbf{x}$ to represent the vector, and capital and bold letter $\mathbf{A}$ to denote the matrix. The element of a vector $\mathbf{x}$ is denoted by a scalar $x_i$, and the element of a matrix $\mathbf{A}$ is denoted by a scalar $A_{ij}$. 
We use $\mathbf{0}$ to represent a vector with each elements equal to zero. Let $\mathbf{x}^T$ and $\mathbf{A}^T$  to denote the transpose of a vector and a matrix, respectively. Let $\langle\cdot,\cdot\rangle$ represent the standard inner product, and let $\|\cdot\|$ denote the $l_2$ norm (the Euclidean norm of a vector or the spectral norm of a matrix). Let matrix norm  $\|\mathbf{x}\|_{\mathbf{M}}=\mathbf{x}^T\mathbf{M}\mathbf{x}$, where $\mathbf{M}$ is a positive semidefinite matrix. We use $\lambda_{\min}(\mathbf{A})$ and $\lambda_{\max}(\mathbf{A})$ to represent the smallest and largest eigenvalues of a symmetric matrix $\mathbf{A}$. The spectral norm of a matrix $\mathbf{A}$ is then given by $\|\mathbf{A}\|=\lambda_{\max}(\mathbf{A}^T\mathbf{A})^{\frac{1}{2}}$. One basic inequality regarding the spectral norm is $\|\mathbf{Ax}\|\leq\|\mathbf{A}\|\|\mathbf{x}\|$.

\begin{definition} \emph{(subdifferential)}
	The subdifferential $\partial f(\mathbf{x})$ of a convex function $f:\mathbb{R}^n\rightarrow \mathbb{R}$ at $\mathbf{x}$ is the set of all subgradients.
	\begin{equation*}
	\partial f(\mathbf{x})=\{\mathbf{g}\in\mathbb{R}^n|\mathbf{g}^T(\mathbf{y}-\mathbf{x})\leq f(\mathbf{y})-f(\mathbf{x}),\forall \mathbf{y}\in\text{dom}(f)\}.
	\end{equation*}
\end{definition}
The definition of subgradients is a generalization of the basic inequality from differentiable convex function to the non-differentiable function. For example, the indicator function over a convex set $I_{\mathcal{C}}(\mathbf{x})=0, \mathbf{x}\in \mathcal{C}$ and $I_{\mathcal{C}}(\mathbf{x})=\infty, \mathbf{x}\notin \mathcal{C}$, is a convex and non-differentiable function. The subdifferential $\partial I_{\mathcal{C}}(\mathbf{x})$ is the classical normal cone $N_{\mathcal{C}}(\mathbf{x})=\{\mathbf{g}|\mathbf{g}^T(\mathbf{y}-\mathbf{x})\leq 0,\forall \mathbf{y}\in \mathcal{C}\}.$

\begin{definition} \emph{(Convex function)}
	A function $f:\mathbb{R}^n\rightarrow \mathbb{R}$ is called  convex with modulus $v\geq 0$ if for all $\mathbf{x},\mathbf{y}\in \text{dom}(f)$ and $\mathbf{g}\in\partial f(\mathbf{x})$, it satisfies
	\begin{equation*}
	f(\mathbf{y})\geq f(\mathbf{x})+\mathbf{g}^T(\mathbf{y}-\mathbf{x})+\frac{v}{2}\|\mathbf{y}-\mathbf{x}\|^2.
	\end{equation*}
\end{definition}
As a consequence of the above definition, we have the following inequality, which will be used in our theoretical development. For arbitrary $ \mathbf{x},\mathbf{y}\in \text{dom}(f),$
\begin{equation}\label{eq:convexineq}
\langle\mathbf{g}_x-\mathbf{g}_y, \mathbf{x}-\mathbf{y}\rangle\geq v\|\mathbf{x}-\mathbf{y}\|^2,  \mathbf{g}_x\in\partial f(\mathbf{x}), \mathbf{g}_y\in\partial f(\mathbf{y}).
\end{equation}
Note that the strictly convex function refers to that modulus $v>0$.
\begin{definition} \emph{(Moreau-Yosida proximal mapping)}
	The proximal mapping of a closed and convex function $f:\mathbb{R}^n\rightarrow \mathbb{R}$ is defined as
	\begin{equation*}
	\textbf{Pr}_{f}(\mathbf{y})=\arg\min\limits_{\mathbf{x}} f(\mathbf{x})+\frac{1}{2}\|\mathbf{x}-\mathbf{y}\|^2.
	\end{equation*}
\end{definition}
If the function $f$ is the indicator function over a closed and convex set $\mathcal{C}$, then $\textbf{Pr}_f(\cdot)=\Pi_{\mathcal{\mathcal{C}}}(\cdot)$ is the metric projection operator over $\mathcal{C}$. For simplicity, we use $[\cdot]_+$ to denote $\Pi_{\mathcal{C}}(\cdot)$ when $\mathcal{C}$ is the positive orthant $[0,+\infty)^n$. One important property of Moreau-Yosida proximal mapping is non-expansiveness, which can be interpreted as the globally Lipschitz continuous with modulus one.
\begin{equation*}
\|\textbf{Pr}_{f}(\mathbf{x})-\textbf{Pr}_{f}(\mathbf{y})\|\leq\|\mathbf{x}-\mathbf{y}\|, \forall \mathbf{x},\mathbf{y}.
\end{equation*}

\subsection{Rationale behind the Algorithm Design}

Algorithm~\ref{alg:frame} is inspired by an inexact Uzawa method in Alternating Directional Method of Multiplier (ADMM). For the sake of brevity, we will use the following vector notation in the rest of the paper. The node-arc incidence matrix $\mathbf{A}^d\in\mathbb{R}^{(N-1)\times L}$ is defined as
\begin{equation*}
\mathbf{A}^d_{nl}=\left \{
\begin{array}{ll}
1, \quad & \text{if } n=\text{Tx}(l)\\
-1,\quad & \text{if } n=\text{Rx}(l) \\
0,\quad & \text{otherwise}
\end{array}
\right.,\forall n\in\mathcal{N}\backslash\{d\}, l\in\mathcal{L}.
\end{equation*}
The matrix $\mathbf{B}^d\in\mathbb{R}^{(N-1)\times F}$ is defined as
\begin{equation*}
\mathbf{B}^d_{nf}=\left \{
\begin{array}{ll}
-1,\quad & \text{if } n=s_f, d=d_f \\
0,\quad & \text{otherwise}
\end{array}
\right.,\forall n\in\mathcal{N}\backslash\{d\}, f\in\mathcal{F}.
\end{equation*}
Define matrix $\mathbf{A}\in\mathbb{R}^{D(N-1)\times DL}$ and $\mathbf{B}\in\mathbb{R}^{D(N-1)\times F}$ as 
\begin{equation*}
\mathbf{A}=\text{diag}\{\mathbf{A}^1,\mathbf{A}^2,\ldots,\mathbf{A}^D\}=
\begin{bmatrix}
\mathbf{A}^1& \cdots &\mathbf{0}  \\
\vdots& \ddots & \vdots\\
\mathbf{0} &\cdots & \mathbf{A}^D
\end{bmatrix}
,
\mathbf{B}=\begin{bmatrix}
\mathbf{B}^1\\
\vdots\\
\mathbf{B}^D
\end{bmatrix}.
\end{equation*}
We denote the objective function $f(\mathbf{x})=U(\mathbf{x})+h(\mathbf{x})$, where the function $U(\mathbf{x})=-\sum_{f\in\mathcal{F}}U_f(x_f)$ and the indicator function $h(\mathbf{x})$ is defined as
\begin{equation*}
h(\mathbf{x})=\left \{
\begin{array}{ll}
0,\quad & \text{if } m_f\leq x_f\leq M_f, \forall f\in\mathcal{F}\\
+\infty,\quad & \text{otherwise}
\end{array}
\right..
\end{equation*}
Let the indicator function $g(\mathbf{r})$ represent the capacity constraints of link rate vector.
\begin{equation*}
g(\mathbf{r})=\left \{
\begin{array}{ll}
0,\quad & \text{if } \left[\sum_{d}r_l^d\right]\in\mathcal{C},r_l^d\geq 0,\forall l,d\\
+\infty,\quad & \text{otherwise}
\end{array}
\right..
\end{equation*}

Based on the above notation, we can reformulate the JCRS problem (\ref{eq:model}) as the following equivalent form. 
\begin{align}
&\min\limits_{\mathbf{x},\mathbf{r}}\quad f(\mathbf{x})+g(\mathbf{r})\label{eq:ADMMmodel}\\
& \begin{array}{r@{\quad}l@{}l@{\quad}l}
s.t. & \mathbf{Bx}+\mathbf{Ar}=\mathbf{0}.\notag
\end{array} 
\end{align}

Note that optimization of this form contains a separable objective function and a separable constraint between injection rate vector $\mathbf{x}$ and the link rate vector $\mathbf{r}$. Therefore, it inspires us to adopt the Alternating Directional Method of Multiplier (ADMM) to split the decision variables $\mathbf{x}$ and $\mathbf{r}$, which results in a nice layered  structure during the operation of the algorithm. Formally, the Augmented Lagrangian function of problem (\ref{eq:ADMMmodel}) is defined as
\begin{equation}\label{eq:augfun}
L(\mathbf{x},\mathbf{r},\boldsymbol{\lambda})= f(\mathbf{x})+g(\mathbf{r})+\frac{\rho}{2}\|\mathbf{Bx}+\mathbf{Ar}-\boldsymbol{\lambda}/\rho\|^2,
\end{equation}
where $\rho$ is a pre-defined penalty parameter, $\boldsymbol{\lambda}$ is the Lagrangian multiplier. Then the ADMM optimizes the Augmented Lagrangian function $L(\mathbf{x},\mathbf{r},\boldsymbol{\lambda})$ in a Gauss-Seidel fashion. In each time slot $t$, go through the following three steps.
\begin{enumerate}
	\item Primal update:  $\mathbf{r}[t]=\argmin\limits_{\mathbf{r}}L(\mathbf{x}[t-1],\mathbf{r},\boldsymbol{\lambda}[t-1])$.
	\item Primal update:  $\mathbf{x}[t]=\argmin\limits_{\mathbf{x}}L(\mathbf{x},\mathbf{r}[t],\boldsymbol{\lambda}[t-1])$.
	\item Dual update:  $\boldsymbol{\lambda}[t]=\boldsymbol{\lambda}[t-1]-\tau\rho(\mathbf{Bx}[t]+\mathbf{Ar}[t])$.
\end{enumerate}
Based on the definition of the matrix $\mathbf{A}$ and $\mathbf{B}$, it is clear that the third step is the virtual queue update (\ref{eq:virtualQ}) in the Algorithm~\ref{alg:frame}. We then show that the second step is indeed the congestion control component in Algorithm~\ref{alg:frame}. We first omit the constant term $g(\mathbf{r}[t])$ and write it as
\begin{equation*}
\mathbf{x}[t]=\argmin\limits_{\mathbf{x}} f(\mathbf{x})+\frac{\rho}{2}\|\mathbf{Bx}+\mathbf{Ar}[t]-\boldsymbol{\lambda}[t-1]/\rho\|^2.
\end{equation*}
Transforming the indicator function $h(\mathbf{x})$ in $f(\mathbf{x})$ into the box constraints, we have
\begin{equation*}
\mathbf{x}[t]=\argmax\limits_{\mathbf{m}\leq\mathbf{x}\leq\mathbf{M}}\sum_{f\in\mathcal{F}}U_f(x_f)-\frac{\rho}{2}\|\mathbf{Bx}+\mathbf{Ar}[t]-\boldsymbol{\lambda}[t-1]/\rho\|^2.
\end{equation*}
Based on the separability of both objective function and box constraints with respect to the variable $x_f$, we can decompose the original problem into $F$ one-dimensional optimization problems.
\begin{align}
x_f[t] = &\argmax\limits_{x_f\in[m_f,M_f]} U_f(x_f)-\frac{\rho}{2}\left(x_f+\sum\limits_{l\in\mathcal{I}(s_f)} r_{l}^{d_f}[t]-\sum\limits_{l\in\mathcal{O}(s_f)} r_{l}^{d_f}[t]+\lambda^{d_f}_{s_f}[t-1]/\rho\right)^2.\notag
\end{align}
Rearranging the terms by utilizing the virtual queue length update in the time slot $t-1$, we can obtain the congestion control component in Algorithm~\ref{alg:frame}.

The next step is to derive the routing component in Algorithm~\ref{alg:frame}. As discussed before, the challenge in the first primal update step of ADMM is that the quadratic term $\|\mathbf{Bx}[t-1]+\mathbf{Ar}-\boldsymbol{\lambda}[t-1]/\rho\|^2$ in the objective function is non-separable with respect to the decision variable $\mathbf{r}$ due to the non-diagonal structure of the matrix $\mathbf{A}$. The basic idea to overcome this difficulty is to \emph{inexactly solve the $\mathbf{r}-$subproblem}, which is based on minimizing a second-order local approximation of the function $\|\mathbf{Bx}[t-1]+\mathbf{Ar}-\boldsymbol{\lambda}[t-1]/\rho\|^2$ instead of the original one. The approximation of the above function at the point $\mathbf{r}[t-1]$ is given by the Taylor expansion.
\begin{align*}
&\left\|\mathbf{Ar}+\mathbf{Bx}[t-1]-\boldsymbol{\lambda}[t-1]/\rho\right\|^2\notag\\
\approx&\text{ constant}+\langle \mathbf{g}[t-1],\mathbf{r}-\mathbf{r}[t-1]\rangle+\|\mathbf{r}-\mathbf{r}[t-1]\|_{\mathbf{M}}^2,
\end{align*}
where the gradient $\mathbf{g}[t-1]=2\mathbf{A}^T(\mathbf{Ar}[t-1]+\mathbf{Bx}[t-1]-\boldsymbol{\lambda}[t-1]/\rho)$ and the matrix $\mathbf{M}$ is diagonal with $\mathbf{M}=\text{diag}\{\ldots,\beta_{l}^d,$ $\ldots\}$. Then, substituting this local approximation into the first step, we can write it as the following form.
\begin{align}
\mathbf{r}[t]=&\argmin\limits_{\mathbf{r}} g(\mathbf{r})+\rho\langle \mathbf{A}^T(\mathbf{Ar}[t-1]+\mathbf{Bx}[t-1]-\boldsymbol{\lambda}[t-1]/\rho),\mathbf{r}-\mathbf{r}[t-1]\rangle+\frac{\rho}{2}\|\mathbf{r}-\mathbf{r}[t-1]\|_{\mathbf{M}}^2.
\end{align}
Transforming the indicator function $g(\mathbf{r})$ into constraints,  we are ready to obtain the routing component in the Algorithm~\ref{alg:frame}. 

The idea of approximately solving the subproblem in the ADMM has been widely applied in the existing literatures~\cite{zhang2011unified,he2015non}. The method is called the inexact Uzawa method and can be actually recovered by the following equivalent form.
\begin{equation}
\mathbf{r}[t]=\arg\min\limits_{\mathbf{r}}L(\mathbf{x}[t-1],\mathbf{r},\boldsymbol{\lambda}[t-1])+\frac{1}{2}\|\mathbf{r}-\mathbf{r}[t-1]\|_{\mathbf{Q}}^2,
\end{equation}
with matrix $\mathbf{Q}=\rho(\mathbf{M}-\mathbf{A}^T\mathbf{A})$. In the sequel, we will use this simplified form to prove all the theoretical results of Algorithm~\ref{alg:frame}.

\subsection{Convergence Analysis}

In this subsection,  we establish the global convergence of Algorithm~\ref{alg:frame}. We first exploit the structure of matrix $\mathbf{B}$ and write the standard ADMM model (\ref{eq:ADMMmodel}) as the following form.
\begin{align}
&\min\limits_{\mathbf{x},\mathbf{r}}\quad f(\mathbf{x})+g(\mathbf{r})\label{eq:newADMMmodel}\\
& \begin{array}{r@{\quad}l@{}l@{\quad}l}
s.t. & \mathbf{A}_s\mathbf{r}=\mathbf{x},\quad \mathbf{A}_r\mathbf{r}=\mathbf{0},\notag
\end{array} 
\end{align}
where $\mathbf{A}_s$ is a $F\times DL$ dimensional matrix formed by extracting the rows of matrix $\mathbf{A}$ whose index node is a source for one flow.
The matrix $\mathbf{A}_r$ is formed by the rest of rows of the matrix $\mathbf{A}$. Therefore, the first equation $\mathbf{A}_s\mathbf{r}=\mathbf{x}$ in (\ref{eq:newADMMmodel}) denotes the flow conservation law in those source nodes and the second equation  $\mathbf{A}_r\mathbf{r}=\mathbf{0}$ describes the flow conservation law in those intermediate nodes. Let the associated Lagrangian multiplier of constraints $\mathbf{A}_s\mathbf{r}=\mathbf{x}$, $\mathbf{A}_r\mathbf{r}=\mathbf{0}$ be $\boldsymbol{\lambda}_s$, $\boldsymbol{\lambda}_r$, respectively and let $\boldsymbol{\lambda}=[\boldsymbol{\lambda}_s;\boldsymbol{\lambda}_r]$. In the sequel, we write the Assumption~\ref{ass:slater} as the following equivalent form.
\begin{assumption}\emph{(Existence of optimal solution)} There exists a saddle point $(\mathbf{x}^*,\mathbf{r}^*,\boldsymbol{\lambda}^*)$ of the problem (\ref{eq:ADMMmodel}), i.e., optimal primal variables $\mathbf{x}^*,\mathbf{r}^*$ and dual variables $\boldsymbol{\lambda}^*$, satisfying the KKT conditions:
	\begin{align}
	&-\boldsymbol{\lambda}_r^*\in\partial f(\mathbf{x}^*),\label{eq:KKT1}\\
	&\mathbf{A}_s^T\boldsymbol{\lambda}_s^*+\mathbf{A}_r^T\boldsymbol{\lambda}_r^*\in\partial g(\mathbf{r}^*),\label{eq:KKT2}\\
	&\mathbf{A}_s\mathbf{r}^*=\mathbf{x}^*, \mathbf{A}_r\mathbf{r}^*=\mathbf{0}.\label{eq:KKT3}
	\end{align}
\end{assumption}
As discussed before, this assumption is a mild condition and can be guaranteed by various conditions. When this assumption fails to hold, Algorithm~\ref{alg:frame} has either unsolvable or unbounded subproblems or a diverging sequence of $\boldsymbol{\lambda}[t]$.
\begin{lemma}\label{lm:suffdescent} \emph{(Sufficient descent of primal and dual variables)} Assume Assumption~\ref{ass:obj} and~\ref{ass:slater}. 
	If $\tau\in [1,(\sqrt{5}+1)/2)$, there exists an $\alpha,\eta>0$ such that 
	\begin{align}
	V(\mathbf{x}[t-1],\mathbf{r}[t-1],\boldsymbol{\lambda}[t-1])-V(\mathbf{x}[t],\mathbf{r}[t],\boldsymbol{\lambda}[t])\geq&\alpha\left(\left\|\begin{bmatrix}\boldsymbol{\lambda}[t-1]-\boldsymbol{\lambda}[t]\\\mathbf{x}[t-1]-\mathbf{x}[t]\end{bmatrix}\right\|^2+\|\mathbf{r}[t-1]-\mathbf{r}[t]\|^2_{\mathbf{Q}}\right)+
	\notag\\
	&2v\|\mathbf{x}[t]-\mathbf{x}^*\|^2+2v\|\mathbf{x}[t]-\mathbf{x}[t-1]\|^2.\label{eq:suffdescent}
	\end{align}
	The function $V(\mathbf{x}[t],\mathbf{r}[t],\boldsymbol{\lambda}[t])$ is defined as
	\begin{align}
	V(\mathbf{x}[t],\mathbf{r}[t],\boldsymbol{\lambda}[t])=&\frac{1}{\rho\tau}\|\boldsymbol{\lambda}[t]-\boldsymbol{\lambda}^*\|^2+\rho\|\mathbf{x}[t]-\mathbf{x}^*\|^2+\|\mathbf{r}[t]-\mathbf{r}^*\|^2_{\mathbf{Q}}+\frac{\rho}{\eta}\|\mathbf{A}_s\mathbf{r}[t]-\mathbf{x}[t]\|^2.
	\end{align}
	where matrix $\mathbf{Q}=\rho(\mathbf{M}-\mathbf{A}^T\mathbf{A})$, $v$ is the convexity modulus of function $f(\mathbf{x})$ and $(\mathbf{x}^*,\mathbf{r}^*,\boldsymbol{\lambda}^*)$ is one of the saddle points of the problem (\ref{eq:ADMMmodel}).
\end{lemma}
In Lemma~\ref{lm:suffdescent}, the function $V(\mathbf{x}[t],\mathbf{r}[t],\boldsymbol{\lambda}[t])$ describes the distance between the current iterates and the optimal solution set. To guarantee that the function $V(\cdot)$ has sufficient descent, the matrix $\mathbf{M}$ should be chosen such that matrix $\mathbf{Q}$ is positive definite with $\|\mathbf{r}[t]-\mathbf{r}^*\|^2_{\mathbf{Q}}>0$. One simple choice is that each diagonal element of matrix $\mathbf{M}$ satisfies
\begin{equation}
\beta_{m,n}^d>\text{deg}(m)+\text{deg}(n), \forall (m,n)\in\mathcal{L}, \forall d\in\mathcal{D}.
\end{equation}
Then one can see that $\mathbf{Q}$ is a diagonally dominant matrix, thus it is also positive definite by Gershgorin circle theorem. Now we are ready to use the sufficient descent of the function $V(\cdot)$ to establish the global convergence of Algorithm~\ref{alg:frame}.
\begin{theorem}\label{thm:conv}\emph{(Global convergence of Algorithm~\ref{alg:frame})} For any $\tau\in [1,(\sqrt{5}+1)/2)$ and any parameter $\beta_{m,n}^d>\emph{deg}(m)+\emph{deg}(n)$ for all $(m,n)\in\mathcal{L}, d\in\mathcal{D}$, the sequences $(\mathbf{x}[t],\mathbf{r}[t],\boldsymbol{\lambda}[t])$ converges to a saddle point of (\ref{eq:ADMMmodel}), namely, 
	\begin{align}
	&\lim\sup\limits_{t\rightarrow\infty} \|\mathbf{x}[t]-\mathbf{x}^*\|=0,\notag\\
	&\lim\sup\limits_{t\rightarrow\infty} \|\mathbf{r}[t]-\mathbf{r}^*\|=0,\\
	&\lim\sup\limits_{t\rightarrow\infty} \|\boldsymbol{\lambda}[t]-\boldsymbol{\lambda}^*\|=0.\notag
	\end{align}
\end{theorem}
Note that the convergence of Algorithm~\ref{alg:frame} only requires the concavity of the utility function without the assumption of smoothness and strictly concavity ($v$ could be zero). The existing theoretical analysis of two-block ADMM~\cite{he2015non} has shown that the algorithm converges at a globally sub-linear rate, i.e., $O(1/\epsilon)$, when both function $f$ and $g$ are proper closed convex. Clearly, our definition of function $f$ and $g$ satisfy this condition and Algorithm~\ref{alg:frame} converges in $O(1/\epsilon)$ iterations. However, in the next subsection, we will present a surprising result that, when the utility function is strictly concave ($v$ is positive), the Algorithm~\ref{alg:frame} actually converges globally and linearly, which requires only  $O(\log(1/\epsilon))$ iterations to achieve an $\epsilon-$accurate solution.

\subsection{Linear Convergence Rate Analysis}

Based on the result in Lemma~\ref{lm:suffdescent}, we have an inequality of the form, for arbitrary $t\geq 1$,
\begin{equation*}
V(\mathbf{x}[t-1],\mathbf{r}[t-1],\boldsymbol{\lambda}[t-1])-V(\mathbf{x}[t],\mathbf{r}[t],\boldsymbol{\lambda}[t])\geq C.
\end{equation*}
To establish the global and linear convergence rate of Algorithm~\ref{alg:frame}, it is sufficient to show that, there exists constant $\gamma>0$ such that
\begin{equation}\label{eq:theotherside}
C\geq \gamma V(\mathbf{x}[t],\mathbf{r}[t],\boldsymbol{\lambda}[t]), \forall t\geq 1.
\end{equation}
The function $V(\mathbf{x}[t],\mathbf{r}[t],\boldsymbol{\lambda}[t])$ contains the terms including $\|\mathbf{r}[t]-\mathbf{r}^*\|^2_{\mathbf{Q}}$ and $\|\boldsymbol{\lambda}[t]-\boldsymbol{\lambda}^*\|^2$, but the lower bound $C$ only contains the terms like $\|\mathbf{r}[t]-\mathbf{r}[t-1]\|^2_{\mathbf{Q}}$. Therefore, the challenge is how to bound the terms $\|\boldsymbol{\lambda}[t]-\boldsymbol{\lambda}^*\|^2$ and $\|\mathbf{r}[t]-\mathbf{r}^*\|^2_{\mathbf{Q}}$ using the existing terms in the lower bound $C$.  In the existing works of theoretical ADMM~\cite{deng2016global}, they assume that the matrix $\mathbf{B}$ is of full row rank and matrix $\mathbf{A}$ is of full column rank, and utilize this assumption to upper bound $\|\boldsymbol{\lambda}[t]-\boldsymbol{\lambda}^*\|^2$ and $\|\mathbf{r}[t]-\mathbf{r}^*\|^2_{\mathbf{Q}}$ by existing terms in $C$. However, in our problem, both matrix $\mathbf{B}$ and $\mathbf{A}$ do not satisfy this assumption (matrix $\mathbf{B}$ has several all-zero rows, i.e., those nodes do not contain sources; the number of rows of matrix $\mathbf{A}$ is less than the number of columns). In the sequel, we provide a completely new theoretical path to overcome this technical challenge. We first introduce some basics in the variational analysis. 

\begin{definition} \emph{(Calmness~\cite{rockafellar1998variational})}
	Define the multi-valued mapping $F:\mathbb{R}^n\rightarrow\mathbb{R}^m$. We say that $F$ is calm at $\mathbf{x}_0$ if there exists a neighborhood $U$ of $\mathbf{x}_0$ and a constant $\kappa_0>0$ such that 
	\begin{equation}
	F(\mathbf{x})\subseteq F(\mathbf{x}_0)+\kappa_0\|\mathbf{x}-\mathbf{x}_0\|\mathbb{B}_y, \forall \mathbf{x}\in U.
	\end{equation}
	where unit ball $\mathbb{B}_y\triangleq\{\mathbf{y}\in\mathbb{R}^m|\|\mathbf{y}\|\leq 1\}$.
\end{definition}
The calmness property can be regarded as a generalization of Lipschitz continuous property from single-valued function to set-valued mapping. Recall that the set-valued mapping $F$ is piecewise polyhedral if the graph of $F$ is the union of finitely many polyhedral sets. The following Lemma in~\cite{robinson1981some} establishes the calmness of the piecewise polyhedral mapping.

\begin{figure}
	\centering
	\includegraphics[width=0.5\textwidth]{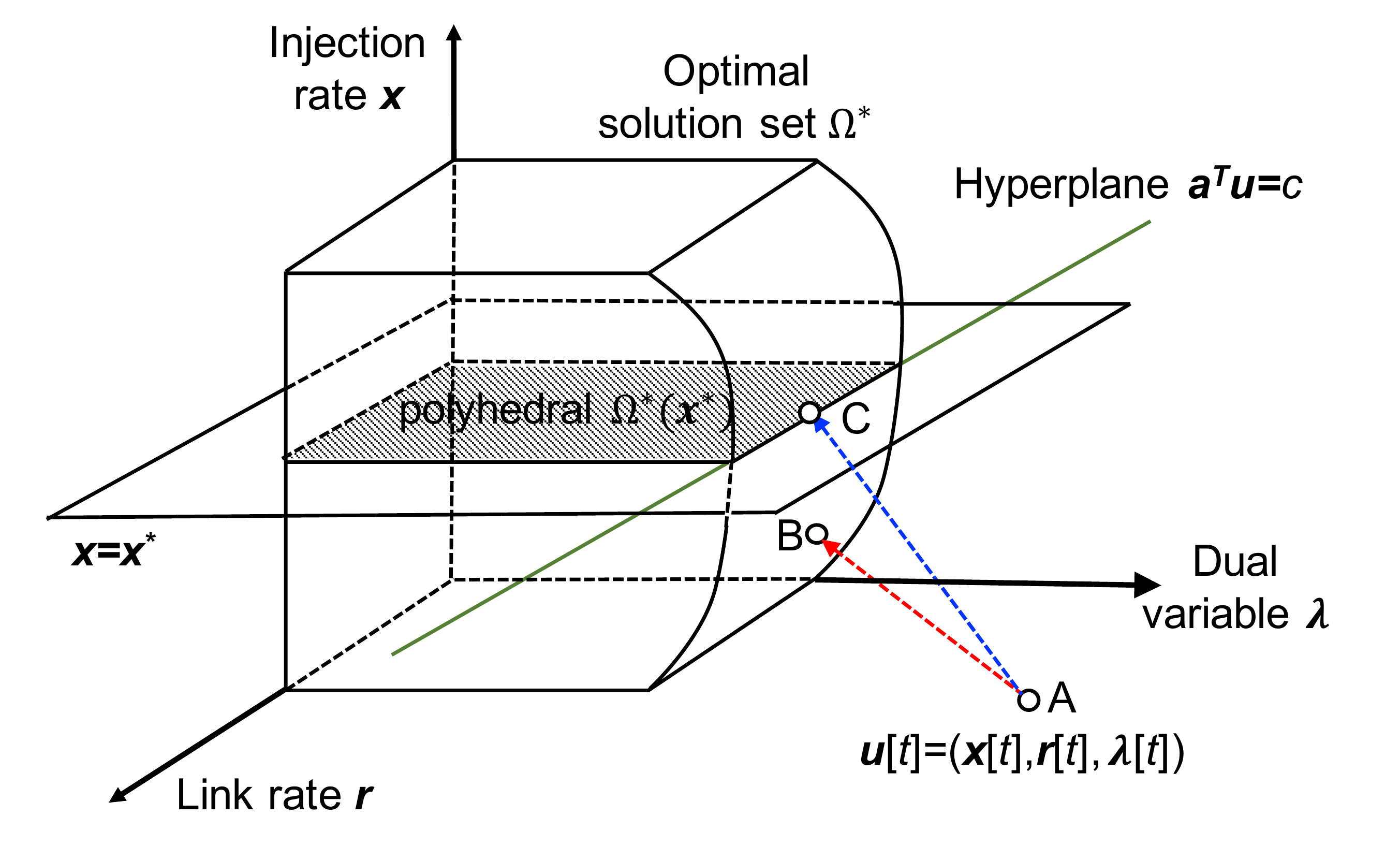}
	\vspace{-0.1in}
	\caption{The distance AB between current iterates and the optimal solution set $\Omega^*$ (non-polyhedral set) is less than the distance AC between current iterates and the set $\Omega^*(\mathbf{x}^*)$ (polyhedral set). The upper bound AC is the distance between a point A and a hyperplane $\mathbf{a}^T\mathbf{u}=c$ that can be implicitly given by $|\mathbf{a}^T\mathbf{u}[t]-b|/\|\mathbf{a}\|=O(|\mathbf{a}^T\mathbf{u}[t]-b|)$ (error bound in the simplest case).} 
	\label{fig:theory}
	\vspace{-0.1in}
\end{figure}

\begin{lemma}\emph{(Calmness of piecewise polyhedral mapping)}\label{lm:calmpfun}
	If the set-valued mapping $F:\mathbb{R}^n\rightarrow\mathbb{R}^m$ is piecewise polyhedral, then $F$ is calm at any $\mathbf{x}_0$ with modulus $\kappa$ independent of choice of $\mathbf{x}_0$.
\end{lemma}

The key technical path to obtain the inequality (\ref{eq:theotherside}) is to utilize the calmness of piecewise polyhedral mapping to establish a global error bound. Then one can apply this error bound to estimate the distance to the optimal solution set, i.e., the terms in the function $V(\mathbf{x}[t],\mathbf{r}[t],\boldsymbol{\lambda}[t])$, by certain constraint violations, which can be further upper bounded by the existing terms in $C$. Denote the solution set of KKT system (\ref{eq:KKT1})-(\ref{eq:KKT3}) by $\Omega^*$. The main difficulty is that the set $\Omega^*$ is non-polyhedron, and one cannot use existing error bound such as Hoffman bound~\cite{hoffman2003approximate} or calmness to estimate the distance to the optimal solution set. However, one important observation is that, the intersection of the optimal solution set $\Omega^*$ and the hyperplane $\mathbf{x}=\mathbf{c}$, given by
\begin{equation}
\Omega^*(\mathbf{c})=\Omega^*\cap\{(\mathbf{x},\mathbf{r},\boldsymbol{\lambda})|\mathbf{x}=\mathbf{c}\},
\end{equation}
is actually the inverse image of a piecewise polyhedral mapping at origin. This result enables us to first upper bound the distance between current iterates $(\mathbf{x}[t],\mathbf{r}[t],\boldsymbol{\lambda}[t])$ and the optimal solution set $\Omega^*$ by the distance to the set $\Omega^*(\mathbf{c})$, then utilize the calmness property to further upper bound above distance by certain constraint violation. 

\begin{lemma}\label{lm:subKKT}
	For arbitrary optimal injection rate vector $\mathbf{x}^*$, define the set-valued mapping $\mathbf{R}_{\mathbf{x}^*}(\mathbf{x},\mathbf{r},\boldsymbol{\lambda})$ as
	\begin{equation}
	\mathbf{R}_{\mathbf{x}^*}(\mathbf{x},\mathbf{r},\boldsymbol{\lambda})=
	\begin{pmatrix}
	\mathbf{x}-\textbf{\emph{Pr}}_{h}(\mathbf{x}-\boldsymbol{\lambda}_s-\nabla U(\mathbf{x}^*))\\ 
	\mathbf{r}-\textbf{\emph{Pr}}_{g}(\mathbf{r}+(\mathbf{A}_s^T\boldsymbol{\lambda}_s+\mathbf{A}_r^T\boldsymbol{\lambda}_r))\\ 
	\mathbf{A}\mathbf{r}+\mathbf{B}\mathbf{x}\\
	\mathbf{x}-\mathbf{x}^*
	\end{pmatrix},
	\end{equation}
	Then, for arbitrary $(\mathbf{x},\mathbf{r},\boldsymbol{\lambda})$, we have $(\mathbf{x},\mathbf{r},\boldsymbol{\lambda})\in\Omega^*(\mathbf{x}^*)$ if and only if $\mathbf{R}_{\mathbf{x}^*}(\mathbf{x},\mathbf{r},\boldsymbol{\lambda})=\mathbf{0}$.
\end{lemma}

Since functions $h(\cdot)$ and $g(\cdot)$ are the indicator functions of the closed and convex sets, the Moreau-Yosida proximal mappings \textbf{Pr}$_h(\cdot)$ and \textbf{Pr}$_g(\cdot)$ are projection mappings onto a convex set and therefore piecewise polyhedral by Proposition 12.30 in~\cite{dontchev2009implicit}. Considering the fact that mappings $\boldsymbol{\lambda}_s+\nabla U(\mathbf{x}^*)$, $\mathbf{A}_s^T\boldsymbol{\lambda}_s+\mathbf{A}_r^T\boldsymbol{\lambda}_r$, $\mathbf{A}\mathbf{r}+\mathbf{B}\mathbf{x}$ and $\mathbf{x}-\mathbf{x}^*$ are affine, the set-valued mapping $\mathbf{R}_{\mathbf{x}^*}(\cdot)$ is therefore piecewise polyhedral, and so is $\mathbf{R}_{\mathbf{x}^*}^{-1}(\cdot)$. Then, from the result of Lemma~\ref{lm:subKKT}, we can regard the subset  $\Omega^*(\mathbf{x}^*)$ as $\mathbf{R}_{\mathbf{x}^*}^{-1}(\mathbf{0})$ and utilize the calmness result in Lemma~\ref{lm:calmpfun} to upper bound the distance between the current iterates and the set $\Omega^*(\mathbf{x}^*)$ by the constraint violation $\|\mathbf{R}_{\mathbf{x}^*}(\mathbf{x}[t],\mathbf{r}[t],\boldsymbol{\lambda}[t])\|$. Formally, we have the following global error bound.

\begin{lemma}\label{lm:errorbound}\emph{(Global error bound)} Assume Assumptions 1 and~\ref{ass:slater}. If $\tau\in[1,(\sqrt{5}+1)/2)$ and parameter $\beta_{m,n}^d>\emph{deg}(m)+\emph{deg}(n)$, then there exists a constant $\kappa>0$ such that the sequence $(\mathbf{x}[t],\mathbf{r}[t],\boldsymbol{\lambda}[t])$ generated by Algorithm~\ref{alg:frame} satisfies
	\begin{equation}\label{eq:errorbound}
	\text{\emph{dist}}^2((\mathbf{x}[t],\mathbf{r}[t],\boldsymbol{\lambda}[t]),\Omega^*)\leq \kappa \|\mathbf{R}_{\mathbf{x}^*}(\mathbf{x}[t],\mathbf{r}[t],\boldsymbol{\lambda}[t])\|^2, t\geq 1,
	\end{equation}
	where $\mathbf{x}^*$ is an arbitrary optimal injection rate vector and the distance function is defined as
	\begin{equation}
	\text{\emph{dist}}^2((\mathbf{x}[t],\mathbf{r}[t],\boldsymbol{\lambda}[t]),\Omega^*)\triangleq \inf\limits_{(\mathbf{x},\mathbf{r},\boldsymbol{\lambda})\in\Omega^*}\left\|\begin{bmatrix}
	\mathbf{x}\\
	\mathbf{r}\\
	\boldsymbol{\lambda}
	\end{bmatrix}-\begin{bmatrix}
	\mathbf{x}[t]\\
	\mathbf{r}[t]\\
	\boldsymbol{\lambda}[t]
	\end{bmatrix}
	\right\|^2,
	\end{equation}
\end{lemma}

We finally upper bound the residual $\|\mathbf{R}_{\mathbf{x}^*}(\mathbf{x}[t],\mathbf{r}[t],\boldsymbol{\lambda}[t])\|$ by the existing terms in lower bound $C$ and combine the results in Lemma~\ref{lm:suffdescent} and Lemma~\ref{lm:errorbound} to establish the global and linear convergence rate in Theorem~\ref{thm:linconv}. The detailed proof can be seen in Appendix~\ref{pf:linconv}. An example of the key proof idea and the global error bound are illustrated in Fig~\ref{fig:theory}.

\subsection{Queue Stability Analysis}

Based on the evolution of physical queue length (\ref{eq:phyQupdate}), we have the following inequality for each queue.
\begin{align}
Q_n^{d}[t]\leq&\left[Q_n^{d}[t-1]-\sum\limits_{l\in\mathcal{O}(n)}r_l^d[t]\right]_ ++\sum\limits_{l\in\mathcal{I}(n)}r_l^d[t]+\sum\limits_{f\in\mathcal{F}}x_f[t]\mathbbm{1}_{\{s_f=n,d_f=d\}}.\label{eq:phyQinq}
\end{align}

In the proof of Theorem~\ref{thm:linconv}, we have shown that the quantity $\mathbf{Bx}[t]+\mathbf{Ar}[t]\leq O(c^t)$, which implies that the change of physical queue length vanishes exponentially. This observation provides a simple path to establish the boundedness of physical queue length. However, it requires the assumption that utility function is strictly concave. In the sequel, we provide a different path, which only assumes the weakly concavity of the utility function. The following technical lemma connects the boundedness of the physical queue length $Q_n^{d}[t]$ and the virtual queue length $\lambda_n^{d}[t]$.
\begin{lemma}\label{lm:vpQ}
	For each destination $d\in\mathcal{D}$ and node $n\in\mathcal{N}\backslash d$, suppose that $\lambda_n^{d}[t]$ and $Q_n^{d}[t]$ evolves by (\ref{eq:virtualQ}) and (\ref{eq:phyQinq}) with initializations $\lambda_n^{d}[t]=Q_n^{d}[t]=0$. If there exists a constant $M>0$ such that $|\lambda_n^{d}[t]|<M,\forall t, d\in\mathcal{D}, n\in\mathcal{N}\backslash d$, then
	\begin{equation}
	Q_n^{d}[t]\leq \frac{2M}{\rho\tau}+B, \forall t, d\in\mathcal{D}, n\in\mathcal{N}\backslash d.
	\end{equation}
	where $B$ is the constant dependent on the largest link capacity.
\end{lemma}
From Theorem~\ref{thm:conv}, we know that the virtual queue length $\boldsymbol{\lambda}[t]$ converges to an optimal dual variable $\boldsymbol{\lambda}^*$ and we can obtain that
\begin{equation*}
|\lambda_n^{d}[t]|\leq\|\boldsymbol{\lambda}[t]\|=\|\boldsymbol{\lambda}[t]-\boldsymbol{\lambda}^*+\boldsymbol{\lambda}^*\|\leq \|\boldsymbol{\lambda}[t]-\boldsymbol{\lambda}^*\|+\|\boldsymbol{\lambda}^*\|.
\end{equation*}
Based on the result in Lemma~\ref{lm:suffdescent}, the function $V(\mathbf{x}[t],\mathbf{r}[t],\boldsymbol{\lambda}[t])$ is monotonically decreasing with respect to $t$. Then we have
\begin{align*}
|\lambda_n^{d}[t]|&\leq \rho\tau V(\mathbf{x}[t],\mathbf{r}[t],\boldsymbol{\lambda}[t]) +\|\boldsymbol{\lambda}^*\|\notag\\
&\leq \rho\tau V(\mathbf{x}[0],\mathbf{r}[0],\boldsymbol{\lambda}[0]) +\|\boldsymbol{\lambda}^*\|\triangleq M, \forall t\geq 1.
\end{align*}
which is a finite constant dependent on the initial distance to the optimal solution set $\Omega^*$. Therefore, combining the result in Lemma~\ref{lm:vpQ}, one can conclude that the physical queue length for each node and destination is finite.

\section{Efficient Subproblem Solver}

In this section, we develop several efficient algorithms to solve the congestion control, routing and scheduling components in Algorithm~\ref{alg:frame}. 

\subsection{Congestion Control}

The congestion control component is an one-dimensional optimization problem, which can be efficiently solved by Newton method or Fibonacci search. Moreover, if the utility function takes a specific form such as the weighted proportional fair utilities, $U_f(x_f)=w_f\log(x_f), x_f>0$, the solution can be obtained in a close-form expression,
\begin{align}
x_f[t]=&\frac{x_f[t-1]}{2}-\frac{z_{s_f}^{d_f}[t]+\Delta r_f[t]}{2\rho}+\sqrt{\frac{w_f}{\rho}+\frac{(z_{s_f}^{d_f}[t]+\Delta r_f[t]-\rho x_f[t-1])^2}{4\rho^2}}.
\end{align}

\subsection{New Backpressure Routing in Wireline Network}

In the wireline network, there exist no interference among different links and the achievable rate region $\mathcal{C}$ is given by the following form~\cite{liu2016joint,yu2017new}.
\begin{equation}
\mathcal{C}=\left\{[r_l^d]\bigg|\sum\limits_{d=1}^D r_l^d\leq C_l, \forall l\right\},
\end{equation}
where $C_l$ is the capacity of link $l$. Then both the objective function and the constraints of problem (\ref{eq:nbpscheduling}) are separable among the rate vectors in different links. Therefore, the link rate $r_l^d[t]$ can be determined in a distributed fashion: for each link $l=(m,n)$, solving the following quadratic program to obtain $[r_{m,n}^d[t],d\in\mathcal{D}]$.
\begin{align}
&\max\limits_{r_{m,n}^d} \sum\limits_{d\in\mathcal{D}}(z_m^d[t]-z_n^d[t])r_{m,n}^d-\frac{\rho\beta_{m,n}}{2}(r_{m,n}^d-r_{m,n}^d[t-1])^2\notag\\
& \begin{array}{r@{\quad}l@{}l@{\quad}l}
s.t. &\sum_{d}r_{m,n}^d\leq C_{m,n}, r_{m,n}^d\geq 0,\forall d.\label{eq:nbprouting}
\end{array} 
\end{align} 
We define this problem as the \textbf{new backpressure routing} problem.  After rearrangement of the terms, it can be formulated as a problem that projects the point $(r_{m,n}^d[t-1]+(z_m^d[t]-z_n^d[t])/\rho\beta_{m,n} )$ onto a simplex defined in (\ref{eq:nbprouting}), which has already been investigated in~\cite{duchi2008efficient}. 
\begin{lemma} \emph{(solution of routing component)} For each link $l=(m,n)\in\mathcal{L}$, the solution of new backpressure routing has the form of $r_{m,n}^d[t]=[r_{m,n}^d[t-1]+(z_m^d[t]-z_n^d[t])/\rho\beta_{m,n}-\theta^*]_+$, where $\theta^*$ can be determined in $O(F\log(F))$ time.
\end{lemma}
The main procedure to solve problem (\ref{eq:nbprouting}) are listed in Algorithm~\ref{alg:nbprouting}. Note that the step 1-4 and step 6-8 have $O(F)$ complexity and hence the overall complexity of Algorithm~\ref{alg:nbprouting} is dominated by the sorting step 5 with complexity $O(F\log(F))$.

\begin{algorithm}[htb]  
	\caption{New backpressure routing algorithm}   
	\label{alg:nbprouting}   
	\renewcommand{\algorithmicrequire}{\textbf{Initialization:}}  
	\renewcommand{\algorithmicensure}{\textbf{Iteration:}} 
	\begin{algorithmic}[1] 
		\STATE Let $x_d=[r_{m,n}^d[t-1]+(z_m^d[t]-z_n^d[t])/\rho\beta_{m,n}]_+, \forall d\in\mathcal{D}$.
		\IF{$\sum_{d=1}^Dx_d\leq C_{m,n}$} 
		\STATE Let $\theta^*=0$ and $r_{m,n}^d[t]=x^d,\forall d\in\mathcal{D}$ and terminate algorithm.
		\ENDIF
		\STATE Sort $\{x_d,d\in\mathcal{D}\}$ in an decreasing order $\pi$ such that $x_{\pi(1)}\geq x_{\pi(2)}\geq\cdots\geq  x_{\pi(D)}$. 
		\STATE Find $p=\max\left\{k\in [D]\big|x_{\pi(k)}-\frac{1}{k}\left(\sum\limits_{d=1}^k x_{\pi(d)}-C_{m,n}\right)>0\right\}$.
		\STATE Let $\theta^*=\frac{1}{p}\left(\sum\limits_{d=1}^p x_{\pi(d)}-C_{m,n}\right)$.
		\STATE Output $r_{m,n}^d[t]=[r_{m,n}^d[t-1]+(z_m^d[t]-z_n^d[t])/\rho\beta_{m,n}-\theta^*]_+,\forall d\in\mathcal{D}$.
	\end{algorithmic}  
\end{algorithm}

\subsection{New Scheduling in Wireless Network}

In the wireless network, different links cannot be simultaneously activated due to the existence of interference. Therefore, in addition to the rate assignment at each link, we need to schedule the link itself. The basic challenge to solve the scheduling component is that the the number of feasible link rate vectors $|\Gamma|$ is possibly exponentially large. For example, in the one-hop node-exclusive model~\cite{lin2004joint}, all the feasible link rate vectors correspond to all the matchings in the graph $\mathcal{G}$, which could be $O(2^L)$ even in the bipartite graph. In the QCA method, the scheduling component is the classical MaxWeight scheduling, and the objective function is linear and such a problem can be reduced to some classical combinatorial problems such as maximum weighted matching. Instead, in our scheduling component (\ref{eq:nbpscheduling}), the objective function is quadratic, and the optimal solution may not belong to the vertex set $\Gamma$ of the convex hull $\mathcal{C}$, which poses a significant challenge in solving this problem. However, utilizing the idea of ellipsoid method, we will show a surprising result that the complexity of solving our new scheduling component (\ref{eq:nbpscheduling}) is equivalent to the complexity of solving the traditional MaxWeight scheduling problem.

Before presenting our main result, we first briefly introduce several concepts and technical tools in geometric algorithms~\cite{grotschel1993geometric} that will be used in the sequel.
\begin{definition}\emph{(Separation oracle)} Let $H$ be a non-empty convex polyhedron in $\mathbb{R}^n$. A separation oracle for $H$ is that, given any $\mathbf{x}\in\mathbb{R}^n$, it either outputs $\mathbf{x}\in H$, and if not, find a hyperplane such that $\mathbf{c}^T\mathbf{x}>\mathbf{c}^T\mathbf{y},\forall \mathbf{y}\in H$.
\end{definition}
\begin{lemma}\label{lm:sepandopt}\emph{(Separation and optimization)} Let $H$ be a non-empty convex polyhedron in $\mathbb{R}^n$ and $f(\cdot)$ be a convex function in $\mathbb{R}^n$. \textbf{If} the separation oracle for $H$ can be solved in \emph{poly}$(n)$ time, then we can compute an $\mathbf{x}$ with $B(\mathbf{x},\delta)\in H$ and $\max_{\mathbf{y}\in H}f(\mathbf{y})-f(\mathbf{x})\geq \delta$ in \emph{poly}$(n,\log(\delta^{-1}))$ time.
\end{lemma}
In this lemma, $B(\mathbf{x},\delta)$ is the ball centering at $\mathbf{x}$ with radius $\delta$, where the $\delta$ is the finite truncation error from irrational number to rational number. 
\begin{theorem} \label{thm:complexityequivalence}
	Assume that the feasible link rate vector $\mathbf{r}^{(i)}\in\mathbb{N}^L, \forall \mathbf{r}^{(i)}\in\Gamma$. There is a \emph{poly}$(L,F)$ time algorithm to compute the new scheduling component (\ref{eq:nbpscheduling}) \textbf{if and only if} there is a \emph{poly}$(L,F)$ time algorithm to compute the MaxWeight scheduling problem (\ref{eq:bpscheduling}).
\end{theorem}
In practice, the link rate always refers to the number of transmitted packets, hence  the integer assumption on the feasible link rate vector is reasonable. Theorem~\ref{thm:complexityequivalence}  shows that \emph{our quadratic scheduling component is not much ``harder'' than the traditional MaxWeight scheduling problem}. Therefore, we can establish the hardness of our new scheduling problem based on all existing complexity results of MaxWeight scheduling. For example, under the node-exclusive interference model, the MaxWeight scheduling is actually a maximum weighted matching problem that can be solved in polynomial time~\cite{lin2004joint}. This result implies that problem (\ref{eq:nbpscheduling}) can also be solved in polynomial time. Another example is the Maximum Weighted K-Valid Matching problem introduced in~\cite{sharma2006complexity} to characterize the multi-hop interference. They show that this problem is NP-hard when we have at least $2-$hop interference, which implies that the problem (\ref{eq:nbpscheduling}) is also NP-hard.

The rest of challenge is the implementation issue incurred by the non-integer solution of (\ref{eq:nbpscheduling}), because the optimal point may not lie in the set of feasible link rate vectors. We next show that this problem can be tackled by connecting the practical time sharing technique and the convex decomposition technique in the combinatorial optimization.
\begin{lemma}\label{lm:sepanddecomp} If there is a \emph{poly}$(L,F)$ time algorithm to compute the MaxWeight scheduling problem (\ref{eq:bpscheduling}), then there is a \emph{poly}$(L,F)$ time algorithm that, given any optimal solution $\mathbf{r}^*$ of (\ref{eq:nbpscheduling}), yields $(L+1)$ feasible rate vectors $\mathbf{r}^{(i)}\in \Gamma$ such that $[\sum_{d}{r_l^d}^*]=\sum_{i=1}^{L+1}\tau_i\mathbf{r}^{(i)}$ and $\sum_{i=1}^{L+1}\tau_i=1,\tau_i\geq 0$.
\end{lemma}
The proof of Lemma~\ref{lm:sepanddecomp} is a straightforward application of Theorem~\ref{thm:complexityequivalence} and the polynomial time reduction from the convex decomposition of point within a polyhedron $H$ to the separation oracle problem~\cite{grotschel1981ellipsoid}. Based on this result, we can first divide the time slots into mini slots and then operate the link rate vector $\mathbf{r}^{(i)}$ in $\tau_i$ fraction of time. For each link $l$, given the link rate vector $r^{(i)}_l$, the specific rate assignment for each $d-$destined packets can be determined by solving a problem same as (\ref{eq:nbprouting}) (only change $C_l$ to $r^{(i)}_l$).

\section{Numerical Analysis}

In this section, we conduct some numerical studies to verify the  theoretical improvements of our proposed method compared with the state-of-arts.

\subsection{Simulation Setup}

We adopt the well-known weighted proportional fair utilities $U_f(x_f)$ $=w_f\log(x_f)$, where the weight $w_f$ of each flow $f$ is randomly generated from a uniform distribution $U(0,1)$. The network typology $\mathcal{G}=(\mathcal{N},\mathcal{L})$ is generated by the classic Erd{\H{o}}s-R{\'e}nyi (ER) random graph model $G(n,p)$, where $n$ is the number of nodes and $p$ is the connected probability between two nodes (we only consider the connected graph). We compare our algorithm with the following three benchmark algorithms.

\textbf{Momentum method:} Here the Momentum method refers to the Heavy-ball algorithm proposed in~\cite{liu2016heavy}.  The existing works~\cite{liu2016heavy,liu2016achieving} have shown that this method produces significantly faster convergence speed and lower queuing delay compared to the traditional QCA method\footnote{Hence, we don't compare our algorithm with QCA.}. 

\textbf{Second-order method:} There exists several versions of second-order algorithms~\cite{wei2013distributed,liu2013distributed,liu2016joint} in solving this problem. We use the one with the fastest convergence speed proposed in~\cite{liu2016joint}. This algorithm has a two-layered iteration structure: (i) each outer iteration corresponds to one Newton step; (ii) a Sherman-Morrison-Woodbury (SMW) based inner iteration to determine the Newton direction. 

\textbf{Proximal method:} We use the one proposed in~\cite{yu2017new}. They have shown superior performance in the queue length reduction and improvement of convergence speed than the QCA method in the wireline network.

We adopt the following two comparison metrics: (i) the relative error of injection rate: $\|\mathbf{x}[t]-\mathbf{x}^*\|/\|\mathbf{x}^*\|$, where the $\mathbf{x}^*$ is obtained approximately by running our method with a strict stopping condition; (ii) total physical queue length of all nodes and all flows: $\sum_{d\in\mathcal{D}}\sum_{n\in\mathcal{N}\backslash d}Q_n^d[t]$. In the simulation, each iteration refers to one communication per node. For our method, momentum method and proximal method, each iteration refers to solving one congestion control and routing component. For the second-order method, each iteration refers to one SMW-based iteration.

\subsection{Wireline Network}

\begin{figure}[!t]
	\includegraphics[width=1\textwidth]{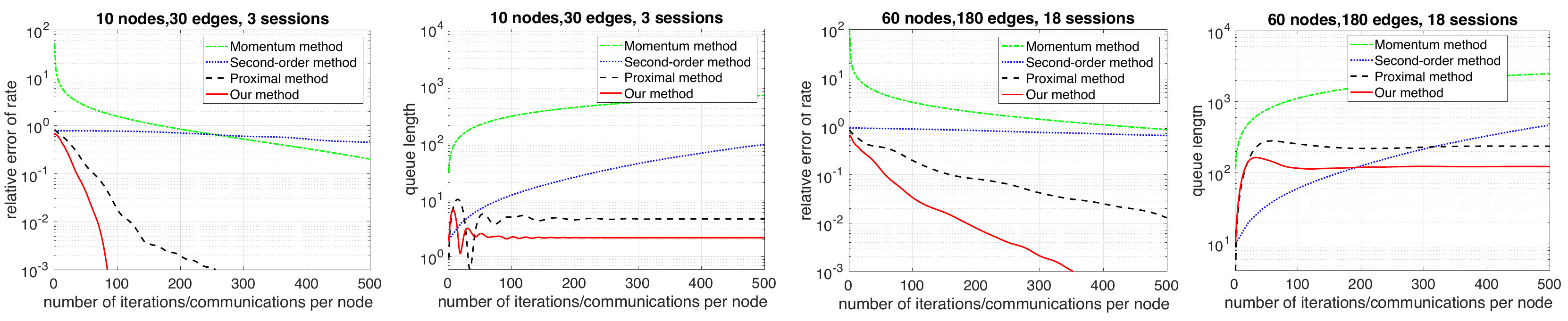}
	\caption{Comparison of Algorithm~\ref{alg:frame} and existing methods in a small-scale and a medium-scale wireline networks.}
	\label{fig:simres1}
	\vspace{-0.1in}
\end{figure}

We first compare our algorithm with above three algorithms in a wireline network with link capacity $C_l$ randomly generated from a uniform distribution $U(0,1)$. As shown in Figure~\ref{fig:simres1}, we plot the relative error of rate and the total physical queue length versus the number of iterations under a small-scale network (10 nodes, 30 edges, 3 sessions) and a medium-scale network (60 nodes, 180 edges, 18 sessions). For the momentum method and second-order method, we choose parameter $K$ and $\mu$ large enough to guarantee the utility optimality gap is less than $0.1\%$. For proximal method, we choose parameter $\alpha_n=(d_n+1)/2$. It can be observed that our proposed algorithm converges at a global and linear rate with bounded physical queue length, which matches our theoretical results. Moreover, it produces the fastest convergence speed and lowest physical queue length among all the existing methods. Although the second-order method has only $40-80$ outer iterations (newton step), it still converges quite slowly due to the large number of inner iterations in computing the Newton direction. Another observation is that our method, proximal method and second-order method gradually increases the injection rates to the optimal point, instead, the momentum method first produces an extremely high injection rate, then gradually decrease it, which leads to a large physical queue length.

\begin{figure}[!t]
	\includegraphics[width=1\textwidth]{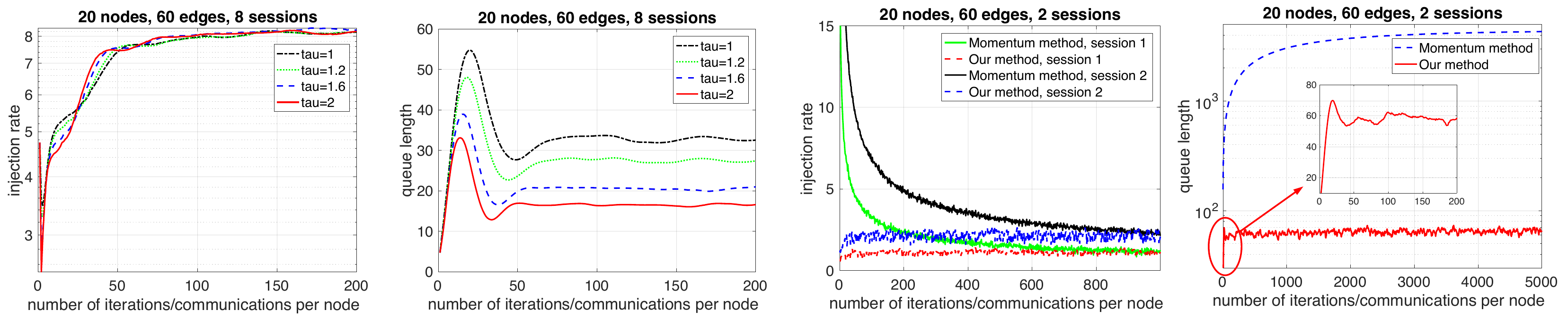}
	\caption{The two left figures shows the impact of parameter $\tau$ on convergence and queue length. The two right figures compare our algorithm and the momentum method for a wireless network with fading channel.}
	\label{fig:simres2}
\end{figure}

\begin{table*}[!t]
	\vspace{-0.1in}
	\caption{Comparison of Convergence Speed and Queue Length per Link}\label{tab:simconv}
	\footnotesize
	\begin{tabular}{c|c|c|c|c|c|c|c|c}
		\hline
		\multirow{2}{*}{Problem size} & \multicolumn{2}{|c|}{Momentum method} & \multicolumn{2}{|c|}{Second-order method} & \multicolumn{2}{|c|}{Proximal method} & \multicolumn{2}{|c}{Our method}\\
		\cline{2-9}
		& $\#$ Iterations & Queue len & $\#$ Iterations & Queue len & $\#$ Iterations & Queue len & $\#$ Iterations & Queue len\\
		\hline
		$(50,150,10)$ & $4658$ & $22.5$ & $ 9600$ & $35.1$ & $369$ & $1.10$ & $207$ &$0.66$\\
		\hline
		$(100,300,20)$ & $9594$ & $82.6$ & $38900$ & $145.2$ & $512$ & $1.94$ & $298$ & $0.83$\\
		\hline
		$(500,1500,100)$ & $>10^5$& $>10^3$ & $>10^5$ & $>10^3$ & $853$ & $8.15$ & $371$ & $3.92$\\
		\hline
		$(1000,3000,200)$ & $>10^5$& $>10^4$ & $>10^5$& $>10^4$ & $1921$ & $15.30$ & $639$ & $6.61$\\
		\hline
		benchmark & $1044$ & $31.2$ & $1510$ & $29.5$ & $102$ & $1.08$ & $82$ & $0.58$\\
		\hline
	\end{tabular}
	\vspace{-0.1in}
\end{table*}

We then investigate the impact of the network size and compare our algorithm with the existing methods in number of iterations and physical queue lengths to obtain solution with a given accuracy. The stopping criterion is that both the relative error of rate $\|\mathbf{x}[t]-\mathbf{x}^*\|/\|\mathbf{x}^*\|$ and the constraint violation $\|\mathbf{Bx}[t]+\mathbf{Ar}[t]\|$ is less than $1\%$. Similarly, we set parameters $K$ and $\mu$ of momentum method and second-order method large enough to guarantee the desired utility optimality gap. To avoid the random noises, we randomly generate $1000$ instances in each problem size and take the average. Besides, we also use a benchmark network in~\cite{yu2017new}, whose optimal solution is known. The results are listed in TABLE~\ref{tab:simconv}. Note that the queue length is normalized by the number of links. It can be observed that our proposed algorithm exhibits a $10-10^3$ order of  improvement of both convergence speed and queue length compared with the momentum method and the second-order method. It also converges $2-3$ times faster and produces $40\%-60\%$ less physical queue length than the proximal method. Moreover, our algorithm has an effect of relieving the curse of dimensionality in the traditional algorithms. For example, when the problem size increases $20$ times (from the first instance to fourth instance), the number of iterations only increases $3$ times. 

\subsection{Impact of Parameter}

We next investigate the impact of parameter $\tau$ on the convergence speed and the queue length of our algorithm. Theorem~\ref{thm:conv} shows that the convergence of our algorithm is guaranteed when $\tau\in [1,(\sqrt{5}+1)/2)$. We test our algorithm in a $20-$nodes $60-$links and $8-$sessions network with $\tau=\{1,1.2,1.6,2.0\}$ and plot the sum of injection rate and queue length versus the number of iterations in Fig.~\ref{fig:simres2}. The basic observation is that when the parameter $\tau$ increases, the convergence speed of our algorithm will slightly increase and the queue length will decrease at an inversely proportional manner, which roughly matches the upper bound of physical queue length provided in Lemma~\ref{lm:vpQ} that $Q_n^{d}[t]\leq \frac{2M}{\rho\tau}+$constant. For example, when $\tau$ increases from $1$ to $2$, the queue length is reduced roughly $40\%$. However, from the simulation, we observe that when $\tau\geq (\sqrt{5}+1)/2$, the algorithm sometimes diverges. Therefore, we suggest a safe value $\tau=1.618$ when using our algorithm.

\subsection{Wireless Network}

From the methods compared above, only the momentum method and our algorithm can be applied to the wireless networks with interference constraints. We compare our algorithm with it in a $20-$nodes $60-$links and $2-$sessions wireless network with quasi-static block fading (channel states vary from one slot to the next but remain
constant in each slot). We plot the injection rate of each session and sum of queue length versus the number of iterations in Fig.~\ref{fig:simres2}. It can be observed that our algorithm converges to the steady state in less than $50$ iterations and the momentum method requires at least $5000$ iterations. Moreover, our algorithm produces only $1\%$ queue length compared to the momentum method.

\section{Discussions}

We now discuss the connection of our algorithm to the existing proximal method and list some follow-up works as well as directions for future research.

\subsection{Connection to Proximal Method}

In the scenario of wireline networks, the existing proximal methods~\cite{lin2006utility,yu2017new} also contain a quadratic regularizer in the congestion control and routing component. Interestingly, we find that this method can be recovered by a kind of proximal linear ADMM with following Jacobi (parallel) updates.
\begin{enumerate}
	\item  $\mathbf{x}[t]=\arg\min\limits_{\mathbf{x}}L(\mathbf{x},\mathbf{r}[t-1],\boldsymbol{\lambda}[t-1])$.
	\item $\mathbf{r}[t]=\arg\min\limits_{\mathbf{r}}L(\mathbf{x}[t-1],\mathbf{r},\boldsymbol{\lambda}[t-1])+\frac{1}{2}\|\mathbf{r}-\mathbf{r}[t]\|^2_{\mathbf{Q}}$.
	\item
	$\boldsymbol{\lambda}[t]=\boldsymbol{\lambda}[t-1]-\rho(\mathbf{Bx}[t]+\mathbf{Ar}[t])$。
\end{enumerate}
The function $L(\cdot)$ is the Augmented Lagrangian function defined in (\ref{eq:augfun}). The matrix $\mathbf{Q}=\rho(\mathbf{M}-\mathbf{A}^T\mathbf{A})$ and matrix  $\mathbf{M}=\text{diag}\{\ldots,\beta^d_l,\ldots\}$. Therefore, we can use the existing analysis in~\cite{deng2017parallel} to establish a stronger theoretical result that the proximal algorithm in~\cite{yu2017new} actually converges in a non-ergodic sublinear rate $o(1/\epsilon)$.

\subsection{ADMM with Acceleration Technique}

There exists some acceleration techniques in the Alternating Directional Method of Multiplier. Similar to the momentum method in the QCA framework, we can introduce some multi-step tricks in the virtual queue length update (\ref{eq:virtualQ}).
\begin{equation*}
\boldsymbol{\lambda}[t]=\boldsymbol{\lambda}[t-1]-\rho(\mathbf{Bx}[t]+\mathbf{Ar}[t])+\alpha[t](\boldsymbol{\lambda}[t-1]-\boldsymbol{\lambda}[t-2]).
\end{equation*}
An open question is that whether this simple trick can provide theoretical improvements in the convergence speed and further reduction of queue length compared with the Algorithm~\ref{alg:frame}.

\subsection{Stochastic Network Optimization}

In the reality, the channel conditions will fluctuate due to the environmental changes (e.g., fading). To accommodate this situation, we assume that there exists a finite set $\mathcal{J}$ of states that channel conditions can be in. Let $\Gamma_j$ denote the set of feasible link rates in state $j$ and $\pi_j$ be the stationary probability of $j$th channel state. We define the following  average capacity region.
\begin{equation*}
\mathcal{C}=\sum\limits_{j\in\mathcal{J}}\pi_j\text{Conv}(\Gamma_j).
\end{equation*}
Then, the problem becomes an optimization problem over this new capacity region. Accordingly, the routing and scheduling components in each time slot $t$ can be modified to an optimization problem over instantaneous region $\mathcal{C}[t]$. Although the numerical results have already exhibited improved performance over existing algorithms, the theoretical performance under this setting is unknown. One possible approach is to utilize some stochastic Alternating Directional Method of Multipliers. However, the challenge is that all existing stochastic ADMMs can only be applied to the smooth stochastic objective function, which is not the case in this problem.

\section{Conclusion}

In this paper, we have proposed a new joint congestion control, routing and scheduling algorithmic framework for distributed network optimization based on an inexact Uzawa method of the Alternating Directional Method of Multiplier. This algorithm offers zero utility optimality gap with finite queue length, the fastest convergence speed to date, i.e., $O(\log(1/\epsilon))$ iterations, among all the existing algorithms. Moreover, the virtual queue-based control provides an extremely low-complexity implementation of this algorithm. These results build a deep connection between the cross-layer decomposition of network optimization and the variable splitting in the multi-block Alternating Directional Method of Multiplier. One important theoretical contribution is that we prove that the ADMM with an inexact Uzawa method converges globally and linearly without requiring the full rank assumption of constraint matrix.

\bibliographystyle{IEEEtran}
\bibliography{sigproc}

\appendix

\subsection{Proof of Lemma~\ref{lm:suffdescent}}
The first step of the Algorithm~\ref{alg:frame} is
\begin{align}
&\mathbf{r}[t]=\arg\min\limits_{\mathbf{r}}L(\mathbf{x}[t-1],\mathbf{r},\boldsymbol{\lambda}[t-1])+\frac{1}{2}\|\mathbf{r}-\mathbf{r}[t-1]\|_{\mathbf{Q}}^2\notag\\
\overset{(a)}{\iff}&\mathbf{r}[t]=\arg\min\limits_{\mathbf{r}}g(\mathbf{r})+\frac{\rho}{2}\left\|\begin{bmatrix}\mathbf{A}_s\mathbf{r}-\mathbf{x}[t-1]-\boldsymbol{\lambda}_s[t-1]/\rho\\\mathbf{A}_r\mathbf{r}-\boldsymbol{\lambda}_r[t-1]/\rho\end{bmatrix}\right\|^2+\frac{1}{2}\|\mathbf{r}-\mathbf{r}[t-1]\|_{\mathbf{Q}}^2\notag\\
\overset{(b)}{\iff}&\mathbf{A}_s^T\left[\boldsymbol{\lambda}_s[t-1]-\rho(\mathbf{A}_s\mathbf{r}[t]-\mathbf{x}[t-1])\right]+\mathbf{A}_r^T\left(\boldsymbol{\lambda}_r[t-1]-\rho\mathbf{A}_r\mathbf{r}[t]\right)+\mathbf{Q}(\mathbf{r}[t-1]-\mathbf{r}[t])\in\partial g(\mathbf{r}[t])\notag\\
\overset{(c)}{\iff}&\mathbf{A}_s^T\left[\bar{\boldsymbol{\lambda}}_s[t]-\rho(\mathbf{x}[t]-\mathbf{x}[t-1])\right]+\mathbf{A}_r^T\bar{\boldsymbol{\lambda}}_r[t]+\mathbf{Q}(\mathbf{r}[t-1]-\mathbf{r}[t])\in\partial g(\mathbf{r}[t]).\label{eq:pf:step1ofADMM}
\end{align}
The above, step (a) utilizes definition of the Augmented Lagrangian function (\ref{eq:augfun}), step (b) is based on the first-order optimality condition, step (c) is based on the following definition of variables $\bar{\boldsymbol{\lambda}}_s[t]$ and $\bar{\boldsymbol{\lambda}}_r[t]$.
\begin{align}
&\bar{\boldsymbol{\lambda}}_s[t]=\boldsymbol{\lambda}_s[t-1]-\rho(\mathbf{A}_s\mathbf{r}[t]-\mathbf{x}[t]),\label{eq:pf:lambda1}\\
&\bar{\boldsymbol{\lambda}}_r[t]=\boldsymbol{\lambda}_r[t-1]-\rho\mathbf{A}_r\mathbf{r}[t].\label{eq:pf:lambda2}
\end{align}
Similarly, based on the the first-order optimality condition and the definition of the variable $\bar{\boldsymbol{\lambda}}_s[t]$ in (\ref{eq:pf:lambda1}), the second step of the Algorithm~\ref{alg:frame} is
\begin{align}
&\mathbf{x}[t]=\arg\min\limits_{\mathbf{x}} f(\mathbf{x})+\frac{\rho}{2}\|\mathbf{A}_s\mathbf{r}[t]-\mathbf{x}-\boldsymbol{\lambda}_s[t-1]/\rho\|^2\notag\\
\iff&-\bar{\boldsymbol{\lambda}}_s[t]\in\partial f(\mathbf{x}[t]).\label{eq:pf:step2ofADMM}
\end{align}
Combining the KKT condition (\ref{eq:KKT2}) that $\mathbf{A}_r^T\boldsymbol{\lambda}_r^*+\mathbf{A}_s^T\boldsymbol{\lambda}_s^*\in\partial g(\mathbf{r}^*)$, the optimality condition (\ref{eq:pf:step1ofADMM}) in the second step of ADMM, we have
\begin{align}
&\langle\mathbf{r}[t]-\mathbf{r}^*,\mathbf{A}_s^T\left[\bar{\boldsymbol{\lambda}}_s[t]-\boldsymbol{\lambda}_s^*-\rho(\mathbf{x}[t]-\mathbf{x}[t-1])\right]+\mathbf{A}_r^T(\bar{\boldsymbol{\lambda}}_r[t]-\boldsymbol{\lambda}_r^*)+\mathbf{Q}(\mathbf{r}[t-1]-\mathbf{r}[t])\rangle\overset{(a)}{\geq} \mathbf{0}.\label{eq:pf:ineq1}
\end{align}
The above, (a) utilizes the fact that function $h(\mathbf{x})$ is convex and the subdifferential of a convex function is a monotone operator, i.e., the inequality (\ref{eq:convexineq}) holds with $v=0$.

Similarly, combining the result of KKT condition (\ref{eq:KKT1}) that $-\boldsymbol{\lambda}_r^*\in\partial f(\mathbf{x}^*)$ and the optimality condition (\ref{eq:pf:step2ofADMM}) in the second step of ADMM, we have
\begin{equation}\label{eq:pf:ineq2}
\langle\mathbf{x}[t]-\mathbf{x}^*,\bar{\boldsymbol{\lambda}}_s[t]-\boldsymbol{\lambda}^*_s[t]\rangle\overset{(a)}{\leq} -v\|\mathbf{x}[t]-\mathbf{x}^*\|^2, v>0.
\end{equation}
The above, step (a) utilizes the fact that function $f(\mathbf{x})$ is convex with modulus $v$ and inequality (\ref{eq:convexineq}).

Then, change the direction of inequality (\ref{eq:pf:ineq2}) and sum it with inequality (\ref{eq:pf:ineq1}), we have
\begin{align}
&\langle\mathbf{x}[t]-\mathbf{x}^*,\boldsymbol{\lambda}^*_s[t]-\bar{\boldsymbol{\lambda}}_s[t]\rangle+\langle\mathbf{r}[t]-\mathbf{r}^*,\mathbf{Q}(\mathbf{r}[t-1]-\mathbf{r}[t])\rangle+\langle\mathbf{r}[t]-\mathbf{r}^*,\mathbf{A}_s^T\left[\bar{\boldsymbol{\lambda}}_s[t]-\boldsymbol{\lambda}_s^*-\rho(\mathbf{x}[t]-\mathbf{x}[t-1])\right]\rangle\notag\\
&+\langle\mathbf{r}[t]-\mathbf{r}^*,\mathbf{A}_r^T(\bar{\boldsymbol{\lambda}}_r[t]-\boldsymbol{\lambda}_r^*)\rangle\geq v\|\mathbf{x}[t]-\mathbf{x}^*\|^2\notag\\
\overset{(a)}{\iff} &\frac{1}{\rho}\langle \boldsymbol{\lambda}_s[t-1]-\bar{\boldsymbol{\lambda}}_s[t],\bar{\boldsymbol{\lambda}}_s[t]-\boldsymbol{\lambda}_s^*\rangle+\rho \langle\mathbf{x}^*-\mathbf{x}[t],\mathbf{x}[t]-\mathbf{x}[t-1]\rangle+\langle\mathbf{r}[t]-\mathbf{r}^*,\mathbf{Q}(\mathbf{r}[t-1]-\mathbf{r}[t])\rangle+\notag\\
&\frac{1}{\rho}\langle \boldsymbol{\lambda}_r[t-1]-\bar{\boldsymbol{\lambda}}_r[t], \bar{\boldsymbol{\lambda}}_r[t]-\boldsymbol{\lambda}_r^*[t]\rangle\geq v\|\mathbf{x}[t]-\mathbf{x}^*\|^2+\langle \boldsymbol{\lambda}_s[t-1]-\bar{\boldsymbol{\lambda}}_s[t],\mathbf{x}[t]-\mathbf{x}[t-1]\rangle\notag\\
\overset{(b)}{\iff}&\frac{1}{\rho}\langle \boldsymbol{\lambda}_s[t-1]-\bar{\boldsymbol{\lambda}}_s[t],\boldsymbol{\lambda}_s[t-1]-\boldsymbol{\lambda}_s^*\rangle+\rho \mathbf{x}[t-1]-\langle\mathbf{x}^*,\mathbf{x}[t-1]-\mathbf{x}[t]\rangle+\langle\mathbf{r}[t-1]-\mathbf{r}^*,\mathbf{Q}(\mathbf{r}[t-1]-\mathbf{r}[t])\rangle+\notag\\
&\frac{1}{\rho}\langle \boldsymbol{\lambda}_r[t-1]-\bar{\boldsymbol{\lambda}}_r[t],\boldsymbol{\lambda}_r[t-1]-\boldsymbol{\lambda}_r^*[t]\rangle\geq \frac{1}{\rho}\left\|\boldsymbol{\lambda}[t-1]-\bar{\boldsymbol{\lambda}}[t]\right\|^2+\rho\|\mathbf{x}[t-1]-\mathbf{x}[t]\|^2+\|\mathbf{r}[t-1]-\mathbf{r}[t]\|^2_{\mathbf{Q}}+\notag\\
&v\|\mathbf{x}[t]-\mathbf{x}^*\|^2+\langle \boldsymbol{\lambda}_s[t-1]-\bar{\boldsymbol{\lambda}}_s[t],\mathbf{x}[t]-\mathbf{x}[t-1]\rangle\notag\\
\overset{(c)}{\iff}&\frac{1}{\rho\tau}\left(\|\boldsymbol{\lambda}[t-1]-\boldsymbol{\lambda}^*\|^2-\|\boldsymbol{\lambda}[t]-\boldsymbol{\lambda}^*\|^2\right)+\rho(\|\mathbf{x}[t-1]-\mathbf{x}^*\|^2-\|\mathbf{x}[t]-\mathbf{x}^*\|^2)+\|\mathbf{r}[t-1]-\mathbf{r}^*\|^2_{\mathbf{Q}}-\|\mathbf{r}[t]-\mathbf{r}^*\|^2_{\mathbf{Q}}\notag\\
&\geq\frac{2-\tau}{\rho}\|\boldsymbol{\lambda}[t-1]-\bar{\boldsymbol{\lambda}}[t]\|^2+\rho\|\mathbf{x}[t-1]-\mathbf{x}[t]\|^2+\|\mathbf{r}[t-1]-\mathbf{r}[t]\|^2_{\mathbf{Q}}+2v\|\mathbf{x}[t]-\mathbf{x}^*\|^2+\notag\\
&2\langle\boldsymbol{\lambda}_s[t-1]-\bar{\boldsymbol{\lambda}}_s[t],\mathbf{x}[t]-\mathbf{x}[t-1]\rangle.\label{eq:pf:suffdescent1}
\end{align}
The above, step (a) rearranges the terms in the original inequality and utilizes the definition of the variable $\bar{\boldsymbol{\lambda}}_s[t], \bar{\boldsymbol{\lambda}}_r[t]$ in (\ref{eq:pf:lambda1}), (\ref{eq:pf:lambda2}) and the KKT condition (\ref{eq:KKT3}). The step (b) rearranges terms by writing $\mathbf{x}^*-\mathbf{x}[t]=\mathbf{x}^*-\mathbf{x}[t-1]+\mathbf{x}[t-1]-\mathbf{x}[t]$ (similarly for variables $\mathbf{r}[t]$ and $\bar{\boldsymbol{\lambda}}[t]$). The step (c) applies the three-point equality of Euclidean norms $\|\mathbf{x}-\mathbf{z}\|^2_{\mathbf{M}}-\|\mathbf{y}-\mathbf{z}\|^2_{\mathbf{M}}=2(\mathbf{x}-\mathbf{z})^T\mathbf{M}(\mathbf{x}-\mathbf{y}) -\|\mathbf{x}-\mathbf{y}\|^2_{\mathbf{M}}$ to the left hand side of the inequality, and utilizes the equation $\boldsymbol{\lambda}[t-1]-\bar{\boldsymbol{\lambda}}[t]=(\boldsymbol{\lambda}[t-1]-\boldsymbol{\lambda}[t])/\tau$. 

The rest is to show that the term $\langle \boldsymbol{\lambda}_s[t-1]-\bar{\boldsymbol{\lambda}}_s[t],\mathbf{x}[t]-\mathbf{x}[t-1]\rangle$ is lower bounded by certain form of other existing terms in (\ref{eq:pf:suffdescent1}) and the primal residual. Applying the equation (\ref{eq:pf:step2ofADMM}) to the time slot $t-1$, we have
\begin{align}
&-\boldsymbol{\lambda}_s[t-2]+\rho(\mathbf{A}_s\mathbf{r}[t-1]-\mathbf{x}[t-1])\in \partial f(\mathbf{x}[t-1]).\notag
\end{align}
Combing this result, equation (\ref{eq:pf:step2ofADMM}) and inequality  (\ref{eq:convexineq}), we have
\begin{align}
&\langle \mathbf{x}[t-1]-\mathbf{x}[t],
-\boldsymbol{\lambda}_s[t-2]+\rho(\mathbf{A}_s\mathbf{r}[t-1]-\mathbf{x}[t-1])+\bar{\boldsymbol{\lambda}}_s[t]\rangle\geq v\|\mathbf{x}[t-1]-\mathbf{x}[t]\|^2\notag\\
\overset{(a)}{\iff} &\langle\mathbf{x}[t]-\mathbf{x}[t-1],
\boldsymbol{\lambda}_s[t-1]-\bar{\boldsymbol{\lambda}}_s[t]\rangle \geq \langle \mathbf{x}[t]-\mathbf{x}[t-1],(1-\tau)\rho(\mathbf{A}_s\mathbf{r}[t-1]-\mathbf{x}[t-1])\rangle+  v\|\mathbf{x}[t-1]-\mathbf{x}[t]\|^2\notag\\
\overset{(b)}{\iff} &\langle \mathbf{x}[t]-\mathbf{x}[t-1],
\boldsymbol{\lambda}_s[t-1]-\bar{\boldsymbol{\lambda}}_s[t]\rangle \geq 
-\frac{\rho}{2\eta}\|\mathbf{A}_s\mathbf{r}[t-1]-\mathbf{x}[t-1]\|^2+\left[v-\frac{(1-\tau)^2\rho\eta}{2}\right]\|\mathbf{x}[t]-\mathbf{x}[t-1]\|^2.\notag
\end{align}
The above, step (a) is based on the virtual queue update $\boldsymbol{\lambda}_s[t-1]=\boldsymbol{\lambda}_s[t-2]-\rho\tau(\mathbf{A}_s\mathbf{r}[t-1]-\mathbf{x}[t-1])$, step (b) utilizes the following inequality.
\begin{align}
&\langle \sqrt{\rho\eta}(1-\tau)(\mathbf{x}[t-1]-\mathbf{x}[t]),\sqrt{\frac{\rho}{\eta}}(\mathbf{A}_s\mathbf{r}[t-1]-\mathbf{x}[t-1])\rangle\leq\notag\\
&\frac{\rho\eta(1-\tau)^2}{2}\|\mathbf{x}[t-1]-\mathbf{x}[t]\|^2+\frac{\rho}{2\eta}\|\mathbf{A}_s\mathbf{r}[t-1]-\mathbf{x}[t-1]\|^2,\notag
\end{align}
where $\eta>1$ is an arbitrary constant. Substituting the above inequality into (\ref{eq:pf:suffdescent1}), we can finally obtain
\begin{align}
&V(\mathbf{x}[t-1],\mathbf{r}[t-1],\boldsymbol{\lambda}[t-1])-V(\mathbf{x}[t],\mathbf{r}[t],\boldsymbol{\lambda}[t])\geq\rho\left(2-\tau-\frac{1}{\eta}\right)\|\mathbf{A}_s\mathbf{r}[t]-\mathbf{x}[t]\|^2+\frac{2-\tau}{\rho}\|\boldsymbol{\lambda}_r[t-1]-\bar{\boldsymbol{\lambda}}_r[t]\|^2+\notag\\
&\rho\left[1-\eta(1-\tau)^2\right]\|\mathbf{x}[t]-\mathbf{x}[t-1]\|^2+\|\mathbf{r}[t-1]-\mathbf{r}[t]\|^2_{\mathbf{Q}}+2v\|\mathbf{x}[t]-\mathbf{x}^*\|^2+2v\|\mathbf{x}[t]-\mathbf{x}[t-1]\|^2.\notag
\end{align}
The existence of $\alpha>0$ can be guaranteed by  $2-\tau-\frac{1}{\eta}> 0$ and $1-\eta(1-\tau)^2> 0$, or, equivalently, $\tau\in[1,(1+\sqrt{5})/2)$. Therefore, the lemma follows.

\section{Proof of Theorem~\ref{thm:conv}}
By Lemma~\ref{lm:suffdescent}, if the parameter $\tau$ satisfies $\tau\in [1,(\sqrt{5}+1)/2)$, the function $V(\mathbf{x}[t],\mathbf{r}[t],\boldsymbol{\lambda}[t])$ is bounded. Then we have that $\|\boldsymbol{\lambda}[t]-\boldsymbol{\lambda}^*\|$, $\|\mathbf{x}[t]-\mathbf{x}^*\|$ and $\|\mathbf{r}[t]-\mathbf{r}^*\|^2_{\mathbf{Q}}$ are bounded, which implies that sequence $\boldsymbol{\lambda}[t]$ and $\mathbf{x}[t]$ are bounded. Based on the choice of parameter $\beta_{m,n}^d>\text{deg}(m)+\text{deg}(n)$, the matrix $\mathbf{Q}$ is positive definite, thus the sequence $\mathbf{r}[t]$ is also bounded. Being bounded, these sequences have the converging subsequences such that
\begin{equation*}
\lim\limits_{i\rightarrow\infty} \mathbf{x}[t_i]=\hat{\mathbf{x}},	\lim\limits_{i\rightarrow\infty} \mathbf{r}[t_i]=\hat{\mathbf{r}},	\lim\limits_{i\rightarrow\infty} \boldsymbol{\lambda}[t_i]=\hat{\boldsymbol{\lambda}}.
\end{equation*}
The function $V(\mathbf{x}[t],\mathbf{r}[t],\boldsymbol{\lambda}[t])$ is monotonically nonincreasing and thus converging. Due to the fact that $\alpha>0$, we have $\lim\sup\|\boldsymbol{\lambda}[t-1]-\boldsymbol{\lambda}[t]\|=0$, and then we have
\begin{equation}\label{eq:pf:kkt3}
\lim\sup\|\mathbf{A}_s\mathbf{r}[t]-\mathbf{x}[t]\|=\lim\sup\|\mathbf{A}_r\mathbf{r}[t]\|=0.
\end{equation}
By passing the limit on (\ref{eq:pf:kkt3}) over subsequences, we have
\begin{equation}\label{eq:pf:kkt4}
\mathbf{A}_s\hat{\mathbf{r}}=\hat{\mathbf{x}}, \mathbf{A}_r\hat{\mathbf{r}}=\mathbf{0}.
\end{equation}
Similarly, we have $\lim\sup\|\mathbf{x}[t-1]-\mathbf{x}[t]\|=\lim\sup\|\mathbf{r}[t-1]-\mathbf{r}[t]\|=0$. Recall the optimality condition (\ref{eq:pf:step1ofADMM}) and (\ref{eq:pf:step2ofADMM}) of first and second step of ADMM, taking limit over the subsequence and applying Theorem 24.4 of~\cite{rockafellar1997convex}, we obtain
\begin{equation}
-\hat{\boldsymbol{\lambda}}_s\in\partial f(\hat{\mathbf{x}}),\text{ and }
\mathbf{A}_s^T\hat{\boldsymbol{\lambda}}_s+\mathbf{A}_r^T\hat{\boldsymbol{\lambda}}_r\in\partial g(\hat{\mathbf{r}}).
\end{equation}
Together with (\ref{eq:pf:kkt4}), $\hat{\mathbf{x}},\hat{\mathbf{r}}, \hat{\boldsymbol{\lambda}}$ satisfy the KKT conditions of problem (\ref{eq:ADMMmodel}). Therefore, the theorem follows.

\section{Proof of Lemma~\ref{lm:subKKT}}
Based on the fact that $f(\mathbf{x}^*)=U(\mathbf{x}^*)$, we have
\begin{align}
&\mathbf{R}_{\mathbf{x}^*}(\mathbf{x},\mathbf{r},\boldsymbol{\lambda})=\mathbf{0}\notag\\
\iff&
\left\{\begin{matrix}
\mathbf{x}-\textbf{Pr}_{h}(\mathbf{x}-\boldsymbol{\lambda}_s-\nabla U(\mathbf{x}^*))=\mathbf{0}\\ 
\mathbf{r}-\textbf{Pr}_{g}(\mathbf{r}+\mathbf{A}_s^T\boldsymbol{\lambda}_s+\mathbf{A}_r^T\boldsymbol{\lambda}_r)=\mathbf{0}\\ 
\mathbf{A}_s\mathbf{r}-\mathbf{x}=\mathbf{0},\mathbf{A}_r\mathbf{r}=\mathbf{0}\\
\mathbf{x}=\mathbf{x}^*
\end{matrix}\right.\notag\\
\overset{(a)}{\iff}&
\left\{\begin{matrix}
-\boldsymbol{\lambda}_s\in\partial h(\mathbf{x})+\nabla U(\mathbf{x}^*)\\ 
\mathbf{A}_s^T\boldsymbol{\lambda}_s+\mathbf{A}_r^T\boldsymbol{\lambda}_r\in\partial g(\mathbf{r})\\
\mathbf{A}_s\mathbf{r}=\mathbf{x}, \mathbf{A}_r\mathbf{r}=\mathbf{0}\\
\mathbf{x}=\mathbf{x}^*
\end{matrix}\right.\notag\\
\overset{(b)}{\iff}&
\left\{\begin{matrix}
-\boldsymbol{\lambda}_s\in\partial f(\mathbf{x})\\ 
\mathbf{A}_s^T\boldsymbol{\lambda}_s+\mathbf{A}_r^T\boldsymbol{\lambda}_r\in\partial g(\mathbf{r})\\
\mathbf{A}_s\mathbf{r}=\mathbf{x}, \mathbf{A}_r\mathbf{r}=\mathbf{0}\\
\mathbf{x}=\mathbf{x}^*
\end{matrix}\right.\notag\\
\iff& (\mathbf{x},\mathbf{r},\boldsymbol{\lambda})\in \Omega^*(\mathbf{x}^*).
\end{align}
The above, step (a) utilizes the definition of proximal mapping and the first-order optimality condition that
\begin{align*}
&\mathbf{x}=\arg\min\limits_{\mathbf{u}} h(\mathbf{u})+\frac{1}{2}\|\mathbf{u}-[\mathbf{x}-\boldsymbol{\lambda}_s-\nabla U(\mathbf{x}^*)]\|^2\notag\\
\iff& \mathbf{0}\in\partial h(\mathbf{x}) +\mathbf{x}-[\mathbf{x}-\boldsymbol{\lambda}_s-\nabla U(\mathbf{x}^*)],
\end{align*}
and 
\begin{align*}
&\mathbf{r}=\arg\min\limits_{\mathbf{u}} g(\mathbf{u})+\frac{1}{2}\|\mathbf{u}-[\mathbf{r}+(\mathbf{A}_s^T\boldsymbol{\lambda}_s+\mathbf{A}_r^T\boldsymbol{\lambda}_r)]\|^2\notag\\
\iff& \mathbf{0}\in\partial g(\mathbf{r}) +\mathbf{r}-[\mathbf{r}+(\mathbf{A}_s^T\boldsymbol{\lambda}_s+\mathbf{A}_r^T\boldsymbol{\lambda}_r)].
\end{align*}
The step (b) is based on the following fact.
\begin{equation*}
\left\{\begin{matrix}
-\boldsymbol{\lambda}_s\in\partial h(\mathbf{x})+\nabla U(\mathbf{x}^*)\\ 
\mathbf{x}=\mathbf{x}^*
\end{matrix}\right. \iff 
\left\{\begin{matrix}
-\boldsymbol{\lambda}_s\in\partial f(\mathbf{x})\\ 
\mathbf{x}=\mathbf{x}^*
\end{matrix}\right..
\end{equation*}
Therefore, the lemma follows.

\section{Proof of Lemma~\ref{lm:errorbound}}
For notational simplicity, let $\mathbf{u}[t]=(\mathbf{x}[t],\mathbf{r}[t],\boldsymbol{\lambda}[t])$. Based on the result of Lemma~\ref{lm:calmpfun}, there exists two constants $\kappa_0$ and $\eta_0$ such that, for all $\mathbf{u}[t]\in \{\mathbf{u}[t]|\mathbf{R}_{\mathbf{x}^*}(\mathbf{u}[t])$ $\leq \eta_0\}$,
\begin{equation}
\text{dist}^2\left(\mathbf{u}[t],\mathbf{R}_{\mathbf{x}^*}^{-1}(\mathbf{0})\right)\leq \kappa_0\|\mathbf{R}_{\mathbf{x}^*}(\mathbf{u}[t])\|^2.
\end{equation}
From Theorem~\ref{thm:conv}, we know that the sequence $\mathbf{u}[t]$ converges to a KKT point $\mathbf{u}^*$ with $\|\mathbf{u}[t]-\mathbf{u}^*\|\leq B_0$ for all $t\geq 1$, where $B_0$ is a finite constant. Then, for $\mathbf{u}[t]$ with $\|\mathbf{R}_{\mathbf{x}^*}(\mathbf{u}[t])\|>\eta_0$, it holds that
\begin{align}
\text{dist}^2\left(\mathbf{u}[t],\mathbf{R}_{\mathbf{x}^*}^{-1}(\mathbf{0})\right)&\leq \|\mathbf{u}[t]-\mathbf{u}^*\|^2\notag\\
&\leq B_0^2\notag\\
&\leq \frac{B_0^2}{\eta_0^2}\|\mathbf{R}_{\mathbf{x}^*}(\mathbf{u}[t])\|^2\notag
\end{align}
Then, let $\kappa=\max\{\kappa_0,B_0^2/\eta_0^2\}$, we have
\begin{equation}
\text{dist}^2\left(\mathbf{u}[t],\mathbf{R}_{\mathbf{x}^*}^{-1}(\mathbf{0})\right)\leq \kappa\|\mathbf{R}_{\mathbf{x}^*}(\mathbf{u}[t])\|^2, \forall t\geq 1.
\end{equation}
Based on the result of Lemma~\ref{lm:subKKT}, the set $\mathbf{R}_{\mathbf{x}^*}^{-1}(\mathbf{0})$ is equivalent to the set $\Omega^*(\mathbf{x}^*)$. Therefore, we have for all $t\geq 1$,
\begin{align}
\text{dist}^2\left(\mathbf{u}[t],\Omega^*\right)&=\inf\limits_{\mathbf{u}\in\Omega^*}\|\mathbf{u}-\mathbf{u}[t]\|^2\notag\\
&\overset{(a)}{\leq}\inf\limits_{\mathbf{u}\in\Omega^*(\mathbf{x^*})}\|\mathbf{u}-\mathbf{u}[t]\|^2\notag\\
&=\text{dist}^2\left(\mathbf{u}[t],\mathbf{R}_{\mathbf{x}^*}^{-1}(\mathbf{0})\right)\notag\\
&\leq  \kappa\|\mathbf{R}_{\mathbf{x}^*}(\mathbf{u}[t])\|^2.
\end{align}
The above, step (a) is based on the definition $\Omega^*(\mathbf{x^*})=\Omega^*\cap\{(\mathbf{x},\mathbf{r},\boldsymbol{\lambda})|\mathbf{x}=\mathbf{x}^*\}$. Therefore, the lemma follows.

\section{Proof of Theorem~\ref{thm:linconv}\label{pf:linconv}}
For notational simplicity, let $\mathbf{u}[t]=(\mathbf{x}[t],\mathbf{r}[t],\boldsymbol{\lambda}[t])$. Based on the result of Lemma~\ref{lm:errorbound}, we have
\begin{align}
&\text{dist}^2\left(\mathbf{u}[t],\Omega^*\right)\leq \kappa\|\mathbf{R}_{\mathbf{x}^*}(\mathbf{u}[t])\|^2=\kappa\left(\|\mathbf{x}[t]-\mathbf{x}^*\|^2+\|\mathbf{x}[t]-\right.\notag\\
&\textbf{Pr}_{h}(\mathbf{x}[t]-\boldsymbol{\lambda}_s[t]-\nabla U(\mathbf{x}^*))\|^2+\|\mathbf{r}[t]-\textbf{Pr}_{g}(\mathbf{r}[t]+\mathbf{A}_s^T\boldsymbol{\lambda}_s[t]+\notag\\
&\mathbf{A}_r^T\boldsymbol{\lambda}_r[t])\|^2+\|\mathbf{A}\mathbf{r}[t]+\mathbf{B}\mathbf{x}[t]\|^2).\label{pf:eq:distineq}
\end{align}
Firstly, the term $\|\mathbf{A}\mathbf{r}[t]+\mathbf{B}\mathbf{x}[t]\|=\|\boldsymbol{\lambda}[t-1]-\boldsymbol{\lambda}[t]\|/\rho\tau$. Secondly, from the Proof of Lemma~\ref{lm:suffdescent}, we have shown that the optimality condition of the first step in Algorithm~\ref{alg:frame} is equivalent to the condition (\ref{eq:pf:step1ofADMM}), which can be further written as
\begin{align}
\mathbf{r}[t]=&\textbf{Pr}_g\left(\mathbf{r}[t]+\mathbf{A}_s^T\left[\bar{\boldsymbol{\lambda}}_s[t]-\rho(\mathbf{x}[t]-\mathbf{x}[t-1])\right]+\mathbf{A}_r^T\bar{\boldsymbol{\lambda}}_r[t]+\mathbf{Q}(\mathbf{r}[t-1]-\mathbf{r}[t])\right).\notag
\end{align}
Then, we have
\begin{align}
&\|\mathbf{r}[t]-\textbf{Pr}_{g}(\mathbf{r}[t]+\mathbf{A}_s^T\boldsymbol{\lambda}_s[t]+\mathbf{A}_r^T\boldsymbol{\lambda}_r[t])\|\notag\\
=&\left\|\textbf{Pr}_g(\mathbf{r}[t]+\mathbf{A}_s^T\left[\bar{\boldsymbol{\lambda}}_s[t]-\rho(\mathbf{x}[t]-\mathbf{x}[t-1])\right]+\mathbf{A}_r^T\bar{\boldsymbol{\lambda}}_r[t]+\mathbf{Q}(\mathbf{r}[t-1]-\mathbf{r}[t]))-\textbf{Pr}_{g}(\mathbf{r}[t]+\mathbf{A}_s^T\boldsymbol{\lambda}_s[t]+\mathbf{A}_r^T\boldsymbol{\lambda}_r[t])\right\|\notag\\
\overset{(a)}{\leq}& \|\mathbf{A}_s^T(\bar{\boldsymbol{\lambda}}_s[t]-\boldsymbol{\lambda}_s[t])+\mathbf{A}_r^T(\bar{\boldsymbol{\lambda}}_r[t]-\boldsymbol{\lambda}_r[t])-\rho\mathbf{A}_s^T(\mathbf{x}[t]-\mathbf{x}[t-1])+\mathbf{Q}(\mathbf{r}[t-1]-\mathbf{r}[t])\|\notag\\
\overset{(b)}{\leq}& \|\mathbf{A}_s^T\|\|\bar{\boldsymbol{\lambda}}_s[t]-\boldsymbol{\lambda}_s[t]\|+\|\mathbf{A}_r^T\|\|\bar{\boldsymbol{\lambda}}_r[t]-\boldsymbol{\lambda}_r[t]\|+\rho\|\mathbf{A}_s^T\|\|\mathbf{x}[t]-\mathbf{x}[t-1]\|+\|\mathbf{Q}\|\|\mathbf{r}[t-1]-\mathbf{r}[t]\|\notag\\
\overset{(c)}{\leq}&(1-\frac{1}{\tau}) \|\mathbf{A}_s^T\|\|\boldsymbol{\lambda}_s[t-1]-\boldsymbol{\lambda}_s[t]\|+\rho\|\mathbf{A}_s^T\|\|\mathbf{x}[t]-\mathbf{x}[t-1]\|+(1-\frac{1}{\tau})\|\mathbf{A}_r^T\|\|\boldsymbol{\lambda}_r[t-1]-\boldsymbol{\lambda}_r[t]\|+\notag\\
&\|\mathbf{Q}\|\|\mathbf{r}[t-1]-\mathbf{r}[t]\|.\notag
\end{align}
The above, step (a) is based on the non-expansiveness of the proximal mapping that $\|\textbf{Pr}_{f}(\mathbf{x})-\textbf{Pr}_{f}(\mathbf{y})\|\leq\|\mathbf{x}-\mathbf{y}\|$, step (b) utilizes the triangle inequality and the matrix norm inequality that $\|\mathbf{Ax}\|\leq\|\mathbf{A}\|\|\mathbf{x}\|$, step (c) is based on the definition of $\bar{\boldsymbol{\lambda}}[t]$ in (\ref{eq:pf:lambda1}) and (\ref{eq:pf:lambda2}) such that $\bar{\boldsymbol{\lambda}}[t]-\boldsymbol{\lambda}[t]=(\tau-1)\rho(\mathbf{Ar}[t]+\mathbf{Bx}[t])=(1-1/\tau)(\boldsymbol{\lambda}[t-1]-\boldsymbol{\lambda}[t])$. Similarly, we have
\begin{align}
\mathbf{x}[t]=\textbf{Pr}_{h}(\mathbf{x}[t]-\bar{\boldsymbol{\lambda}}_s[t]-\nabla U(\mathbf{x}[t])),
\end{align}
and then
\begin{align}
&\|\mathbf{x}[t]-\textbf{Pr}_{h}(\mathbf{x}[t]-\boldsymbol{\lambda}_s[t]-\nabla U(\mathbf{x}^*))\|\notag\\
=&\|\textbf{Pr}_{h}(\mathbf{x}[t]-\bar{\boldsymbol{\lambda}}_s[t]-\nabla U(\mathbf{x}[t]))-\textbf{Pr}_{h}(\mathbf{x}[t]-\boldsymbol{\lambda}_s[t]-\nabla U(\mathbf{x}^*))\|\notag\\
\leq&\|\bar{\boldsymbol{\lambda}}_s[t]-\boldsymbol{\lambda}_s[t]\|+\|\nabla U(\mathbf{x}[t])-\nabla U(\mathbf{x}^*)\|\notag\\
\overset{(d)}{\leq}&(1-\frac{1}{\tau})\|\boldsymbol{\lambda}_s[t-1]-\boldsymbol{\lambda}_s[t]\|+L_u\|\mathbf{x}[t]-\mathbf{x}^*\|.
\end{align}
The above, step (d) is based on the assumption that utility function $U(\cdot)$ has Lipschitz continuous gradient with constant $L_u$. Then, substitute the above inequalities into upper bound (\ref{pf:eq:distineq}) and rearrange the terms, we have
\begin{align}
\text{dist}^2\left(\mathbf{u}[t],\Omega^*\right)&\leq c_1\|\mathbf{x}^*-\mathbf{x}[t]\|^2+c_2\|\boldsymbol{\lambda}[t-1]-\boldsymbol{\lambda}[t]\|^2+c_3\|\mathbf{x}[t]-\mathbf{x}[t-1]\|^2+c_4\|\mathbf{r}[t]-\mathbf{r}[t-1]\|^2\label{eq:pf:errorbound},
\end{align}
where the constant $c_1,c_2,c_3$ and $c_4$ are given by
\begin{align}
&c_1=\kappa(1+2L_u^2),\notag\\
&c_2=(1-\frac{1}{\tau})^2(4\max\{\|\mathbf{A}_s^T\|^2,\|\mathbf{A}_r^T\|^2\}+2)+\frac{1}{\rho^2\tau^2},\notag\\
&c_3=4\rho^2\|\mathbf{A}_s^T\|^2,\notag\\
&c_4=4\|\mathbf{Q}\|^2\notag.
\end{align}
Note that the constants $2$ and $4$ in coefficients $c_i$ derive from the Cauchy-Schwartz inequality. For all $t\geq 1$, define
\begin{equation*}
(\overline{\mathbf{x}}_t,\overline{\mathbf{r}}_t,\overline{\boldsymbol{\lambda}}_t)=\arg\min\limits_{(\mathbf{x},\mathbf{r},\boldsymbol{\lambda})\in\Omega^*} \|\mathbf{x}-\mathbf{x}[t]\|^2+\|\mathbf{r}-\mathbf{r}[t]\|^2+\|\boldsymbol{\lambda}-\boldsymbol{\lambda}[t]\|^2.
\end{equation*}
Then we have
\begin{equation}
\text{dist}^2\left(\mathbf{u}[t],\Omega^*\right)=\|\mathbf{x}[t]-\overline{\mathbf{x}}_t\|^2+\|\mathbf{r}[t]-\overline{\mathbf{r}}_t\|^2+\|\boldsymbol{\lambda}[t]-\overline{\boldsymbol{\lambda}}_t\|^2.
\end{equation}
Further, define
\begin{align}
&\mathbf{x}^*_{t}=\arg\min\limits_{(\mathbf{x},\mathbf{r},\boldsymbol{\lambda})\in\Omega^*} \|\mathbf{x}-\mathbf{x}[t]\|,\notag\\
&\mathbf{r}^*_{t}=\arg\min\limits_{(\mathbf{x},\mathbf{r},\boldsymbol{\lambda})\in\Omega^*} \|\mathbf{r}-\mathbf{r}[t]\|,\notag\\
&\boldsymbol{\lambda}_t^*=\arg\min\limits_{(\mathbf{x},\mathbf{r},\boldsymbol{\lambda})\in\Omega^*} \|\boldsymbol{\lambda}-\boldsymbol{\lambda}[t]\|\notag.
\end{align}

Based on the fact that matrix $\mathbf{Q}$ is positive definite, we have $\lambda_{\min}(\mathbf{Q})>0$ and  $\|\mathbf{r}[t]-\mathbf{r}^*\|^2_{\mathbf{Q}}\geq \lambda_{\min}(\mathbf{Q})\|\mathbf{r}[t]-\mathbf{r}^*\|^2$. Then, we can write the inequality (\ref{eq:suffdescent}) in Lemma~\ref{lm:suffdescent} as the following form.
\begin{align}
V(\mathbf{x}[t-1],\mathbf{r}[t-1],\boldsymbol{\lambda}[t-1])-V(\mathbf{x}[t],\mathbf{r}[t],\boldsymbol{\lambda}[t])\geq& c_5\|\boldsymbol{\lambda}[t-1]-\boldsymbol{\lambda}[t]\|+c_5\|\mathbf{x}[t-1]-\mathbf{x}[t]\|^2+c_6\|\mathbf{r}[t]-\mathbf{r}[t-1]\|^2\notag\\
+&c_7\|\mathbf{x}[t]-\mathbf{x}^*\|^2+c_7\|\mathbf{x}[t]-\mathbf{x}[t-1]\|^2,
\end{align}
where the coefficients $c_5,c_6$ and $c_7$ are positive constants. Combining the above inequality with the error bound (\ref{eq:pf:errorbound}), we conclude that there exists a positive constant $\gamma>0$ such that
\begin{align}
&V(\mathbf{x}[t-1],\mathbf{r}[t-1],\boldsymbol{\lambda}[t-1])-V(\mathbf{x}[t],\mathbf{r}[t],\boldsymbol{\lambda}[t])\geq\gamma\left(\frac{1}{\rho\tau}\|\boldsymbol{\lambda}[t]-\overline{\boldsymbol{\lambda}}_t\|^2+\rho\|\mathbf{x}[t]-\overline{\mathbf{x}}_t\|^2+\|\mathbf{r}[t]-\overline{\mathbf{r}}_t\|_{\mathbf{Q}}^2\right.\notag\\
&\left.+\frac{\rho}{\eta}\|\mathbf{A}_s\mathbf{r}[t]-\mathbf{x}[t]\|^2\right).\notag
\end{align}
Let $\mathbf{x}^*=\mathbf{x}^*_{t-1}$, $\mathbf{r}^*=\mathbf{r}^*_{t-1}$ and $\boldsymbol{\lambda}^*=\boldsymbol{\lambda}^*_{t-1}$ in the function $V(\cdot)$ of the above inequality, then we have
\begin{align}
&\frac{1}{\rho\tau}\|\boldsymbol{\lambda}[t-1]-\boldsymbol{\lambda}_{t-1}^*\|^2+\rho\|\mathbf{x}[t-1]-\mathbf{x}_{t-1}^*\|^2+\frac{\rho}{\eta}\|\mathbf{A}_s\mathbf{r}[t-1]-\mathbf{x}[t-1]\|^2+\|\mathbf{r}[t-1]-\mathbf{r}_{t-1}^*\|^2_{\mathbf{Q}}\geq\notag\\
&\left(\frac{1}{\rho\tau}\|\boldsymbol{\lambda}[t]-\boldsymbol{\lambda}_{t-1}^*\|^2+\rho\|\mathbf{x}[t]-\mathbf{x}_{t-1}^*\|^2+\|\mathbf{r}[t]-\mathbf{r}^*_{t-1}\|^2_{\mathbf{Q}}+\frac{\rho}{\eta}\|\mathbf{A}_s\mathbf{r}[t]-\mathbf{x}[t]\|^2\right)+\gamma\left(\frac{1}{\rho\tau}\|\boldsymbol{\lambda}[t]-\overline{\boldsymbol{\lambda}}_t\|^2+\right.\notag\\
&\rho\|\mathbf{x}[t]-\overline{\mathbf{x}}\|^2+\|\mathbf{r}[t]-\overline{\mathbf{r}}_t\|_{\mathbf{Q}}^2+\frac{\rho}{\eta}\|\mathbf{A}_s\mathbf{r}[t]-\mathbf{x}[t]\|^2).\label{eq:pf:rate1}
\end{align}
Based on the definition of sequences $(\overline{\mathbf{x}}_t,\overline{\mathbf{r}}_t,\overline{\boldsymbol{\lambda}}_t)$ and $(\mathbf{x}^*_t,\mathbf{r}^*_t,\boldsymbol{\lambda}^*_t)$, we have
\begin{align}
&\|\mathbf{x}[t]-\mathbf{x}_{t-1}^*\|\geq \|\mathbf{x}[t]-\mathbf{x}_{t}^*\| , \|\mathbf{x}[t]-\overline{\mathbf{x}}\| \geq \|\mathbf{x}[t]-\mathbf{x}_{t}^*\|,\notag\\
&\|\mathbf{r}[t]-\mathbf{r}_{t-1}^*\|\geq \|\mathbf{r}[t]-\mathbf{r}_{t}^*\| , \|\mathbf{r}[t]-\overline{\mathbf{x}}\| \geq \|\mathbf{r}[t]-\mathbf{r}_{t}^*\|,\notag\\
&\|\boldsymbol{\lambda}[t]-\boldsymbol{\lambda}_{t-1}^*\|\geq \|\boldsymbol{\lambda}[t]-\boldsymbol{\lambda}_{t}^*\|, \|\boldsymbol{\lambda}[t]-\overline{\boldsymbol{\lambda}}_t\|\geq \|\boldsymbol{\lambda}[t]-\boldsymbol{\lambda}_{t}^*\|.\label{eq:pf:rate2}
\end{align}
Combining inequality (\ref{eq:pf:rate1}) and (\ref{eq:pf:rate2}) together, we can get the following contraction.
\begin{equation*}
G[t]\leq \frac{1}{1+\gamma}G[t-1], t\geq 1.
\end{equation*}
where $G[t]$ is defined as
\begin{align}
G[t]=&\frac{1}{\rho\tau}\|\boldsymbol{\lambda}[t]-\boldsymbol{\lambda}_{t}^*\|^2+\rho\|\mathbf{x}[t]-\mathbf{x}_{t}^*\|^2+\|\mathbf{r}[t]-\mathbf{r}^*_{t}\|^2_{\mathbf{Q}}+\frac{\rho}{\eta}\|\mathbf{A}_s\mathbf{r}[t]-\mathbf{x}[t]\|^2.\label{eq:pf:Qlinear}
\end{align}
Telescoping the above inequality for all iterations $t$, we arrive that
\begin{align}
&G[t]\leq \left(\frac{1}{1+\gamma}\right)^t D_0,
\end{align}
where $D_0$ is the initial distance to the optimal solution set, 
\begin{align}
D_0=&\frac{1}{\rho\tau}\|\boldsymbol{\lambda}[0]-\boldsymbol{\lambda}_{0}^*\|^2+\rho\|\mathbf{x}[0]-\mathbf{x}_{0}^*\|^2+\|\mathbf{r}[0]-\mathbf{r}^*_{0}\|^2_{\mathbf{Q}}+\frac{\rho}{\eta}\|\mathbf{A}_s\mathbf{r}[0]-\mathbf{x}[0]\|^2.\notag
\end{align}
Therefore, the theorem follows.
\section{Proof of Lemma~\ref{lm:vpQ}}

Define an auxiliary queue $\hat{\lambda}_n^d[t]$ that evolves according to (\ref{eq:virtualQ}). Initializing the auxiliary queue with $\hat{\lambda}_n^d[0]=M+\rho\tau\sum_{l\in\mathcal{O}(n)}\eta_l$, where $\eta_l$ is the upper bound of the capacity of link $l$. Then we can prove by induction that
\begin{equation*}
\hat{\lambda}_n^d[t]=\lambda_n^d[t]+M+\rho\tau\sum\limits_{l\in\mathcal{O}(n)}\eta_l,\forall t,d\in\mathcal{D}, n\in\mathcal{N}\backslash d.
\end{equation*}
Since $\lambda_n^d[t]\geq -M, \forall t,n,d$ by assumption, we have that  
\begin{equation*}
\hat{\lambda}_n^d[t]\geq \rho\tau\sum_{l\in\mathcal{O}(n)}\eta_l,\forall t,d\in\mathcal{D}, n\in\mathcal{N}\backslash d.
\end{equation*}
Then the auxiliary queue $\hat{\lambda}_n^d[t]$ satisfies
\begin{align}
\hat{\lambda}_n^{d}[t]=&\left[\hat{\lambda}_n^{d}[t-1]-\rho\tau\sum\limits_{l\in\mathcal{O}(n)}r_l^d[t]\right]_+ +\rho\tau\sum\limits_{l\in\mathcal{I}(n)}r_l^d[t]+ \rho\tau\sum\limits_{f\in\mathcal{F}}x_f[t]\mathbbm{1}_{\{s_f=n,d_f=d\}}, \forall t,d\in\mathcal{D}, n\in\mathcal{N}\backslash d.\notag
\end{align}
Based on the fact that $\rho\tau>0$, we can rewrite the above updating formula as
\begin{align}
\frac{\hat{\lambda}_n^{d}[t]}{\rho\tau}=&\left[\frac{\hat{\lambda}_n^{d}[t-1]}{\rho\tau}-\sum\limits_{l\in\mathcal{O}(n)}r_l^d[t]\right]_++\sum\limits_{l\in\mathcal{I}(n)}r_l^d[t]+ \sum\limits_{f\in\mathcal{F}}x_f[t]\mathbbm{1}_{\{s_f=n,d_f=d\}}.\notag
\end{align}
We next prove that $Q_n^d[t]\leq\hat{\lambda}_n^{d}[t]/\rho\tau,\forall t\geq 1$ by induction. For $t=0$, we have $Q_n^d[0]=0\leq \hat{\lambda}_n^{d}[0]/\rho\tau$. Suppose that it holds for $k=t-1$, then for $k=t$, we have
\begin{align}
Q_n^{d}[t]\leq&\left[Q_n^{d}[t-1]-\sum\limits_{l\in\mathcal{O}(n)}r_l^d[t]\right]_+ +\sum\limits_{l\in\mathcal{I}(n)}\hat{r}_l^d[t]+\sum\limits_{f\in\mathcal{F}}x_f[t]\mathbbm{1}_{\{s_f=n,d_f=d\}}\notag\\
\leq&\left[\frac{\hat{\lambda}_n^{d}[t-1]}{\rho\tau}-\sum\limits_{l\in\mathcal{O}(n)}r_l^d[t]\right]_+ +\sum\limits_{l\in\mathcal{I}(n)}\hat{r}_l^d[t]+\sum\limits_{f\in\mathcal{F}}x_f[t]\mathbbm{1}_{\{s_f=n,d_f=d\}}\notag\\
=&\frac{\hat{\lambda}_n^{d}[t]}{\rho\tau}.
\end{align}
Finally, since $\hat{\lambda}_n^d[t]=\lambda_n^d[t]+M+\rho\tau\sum_{l\in\mathcal{O}(n)}\eta_l$ and $\lambda_n^d[t]\leq M$, we have
\begin{equation*}
Q_n^{d}[t]\leq\frac{2M}{\rho\tau}+\sum_{l\in\mathcal{O}(n)}\eta_l.
\end{equation*}
Let constant $B=\max_{n\in\mathcal{N}}\sum_{l\in\mathcal{O}(n)}\eta_l.$ Therefore, the lemma follows.

\section{Proof of Theorem~\ref{thm:complexityequivalence}}

Let $\mathcal{P}\in\mathbb{R}^{L(D+1)}$ be a convex polyhedron, defined as
\begin{equation*}
\mathcal{P}=\left\{(\mathbf{y},\mathbf{r})\bigg|\mathbf{y}\in\mathcal{C},y_l=\sum\limits_{d=1}^D r_{l}^d, \text{ and } r_{l}^d\geq 0,\forall l\in\mathcal{L},d\in\mathcal{D} \right\}.
\end{equation*}
Formally, we define following two problems. The first one is the scheduling component in Algorithm~\ref{alg:frame}.
\begin{definition}\emph{(New scheduling problem)} Given arbitrary weights $\mathbf{a}\in\mathbb{R}^{DL}$, $\mathbf{b}\in\mathbb{R}^{DL}$ and $\mathbf{c}\in\mathbb{R}^L$, output an $(\mathbf{r}^*,\mathbf{y}^*)$ such that, for arbitrary $(\mathbf{r},\mathbf{y})\in\mathcal{P}$,
	\begin{equation}\label{eq:pf:newmaxweight}
	\sum\limits_{l=1}^{L}\sum\limits_{d=1}^{D}a_{l}^d {r_l^d}^* - c_l({r_l^d}^*-b_l^d)^2\geq -\delta+ \sum\limits_{i=1}^{L}\sum\limits_{j=1}^{D}a_{l}^d r_l^d - c_i(r_l^d-b_l^d)^2,
	\end{equation}	
	and $B((\mathbf{r}^*,\mathbf{y}^*),\delta)\in\mathcal{P}$.
\end{definition}
\begin{definition}\emph{(MaxWeight scheduling)} Given arbitrary weights $\mathbf{w}\in\mathbb{Z}^D$, output an $\mathbf{r}^*\in\mathcal{C}$ such that
	\begin{equation}\label{eq:pf:maxweight}
	\mathbf{w}^T\mathbf{r}^*\geq \mathbf{w}^T\mathbf{r}, \forall \mathbf{r}\in\mathcal{C} \text{ and } r_{l}\geq 0.
	\end{equation}	
\end{definition}
We can observe that the above defined problem is actually equivalent to the original MaxWeight scheduling problem (\ref{eq:bpscheduling}) based on the fact that
\begin{align}
&\max\limits_{r_{l}^d}\sum\limits_{l=1}^L \sum\limits_{d\in\mathcal{D}}(Q_m^d[t]-Q_n^d[t])r_{l}^d,\text{ s.t. }\left[\sum_{d}r_l^d\right]\in\mathcal{C}, r_{l}^d\geq 0.\notag\\
\iff&\max\limits_{\mathbf{r}}\sum\limits_{l=1}^L (Q_m^{d_l}[t]-Q_n^{d_l}[t])r_{l}^{d_l},\text{ s.t. }\mathbf{r}\in\mathcal{C},r_{l}^{d_l}\geq 0,\notag
\end{align}
where $d_l$ is defined as $d_l=\arg\max_{d\in\mathcal{D}}(Q_m^d[t]-Q_n^d[t])$, and the fact that the physical queue length in the QCA method is an integer (number of packets). According to the above definitions, to prove the Theorem~\ref{thm:complexityequivalence}, we need to construct a poly$(L,F)$ time reduction between the above two problems. 

We first prove the ``if'' direction. 

Based on the result in Lemma~\ref{lm:sepandopt},  we know that solving the new scheduling problem in poly$(L,F,\log(\delta^{-1}))$ time if the separation oracle problem for polyhedron $\mathcal{P}$ can be solved in poly$(L,F)$ time. Since the constraints $y_l=\sum_{d=1}^D r_l^d$ and $r_l^d\geq 0$ in $\mathcal{P}$ can be explicitly checked in $O(LF)$ time, then the separation oracle problem for polyhedron $\mathcal{P}$ can be reduced to the separation oracle problem for polyhedron $\mathcal{C}$ by the following procedure: given a separation hyperplane $\mathbf{c^T}\mathbf{y}\geq\mathbf{c}^T\mathbf{y}',\forall \mathbf{y}'\in\mathcal{C}$, construct the hyperplane $\mathbf{c}^T\mathbf{y}+\mathbf{c'}^T\mathbf{r}$ with ${c_{l}^d}'=c_l,\forall l,d$. Then, we have
\begin{align*}
\mathbf{c}^T\mathbf{y}+\mathbf{c'}^T\mathbf{r}&=\mathbf{c}^T\mathbf{y}+\sum\limits_{l=1}^Lc_l\sum\limits_{D=1}^D r_{l}^d=\mathbf{c}^T\mathbf{y}+\mathbf{c}^T\mathbf{y}\\
&\geq \mathbf{c}^T\mathbf{y}'+\mathbf{c}^T\mathbf{y}'\\
&=\mathbf{c}^T\mathbf{y}'+\mathbf{c'}^T\mathbf{r}',\forall (\mathbf{y}',\mathbf{r}')\in\mathcal{P},
\end{align*}
which implies that $\mathbf{c}^T\mathbf{y}+\mathbf{c'}^T\mathbf{r}$ is also a separating hyperplane of polyhedron $\mathcal{P}$. A classic result in the combinatorial optimization due to Gr{\"o}tschel and Lov{\'a}sz~\cite{grotschel1981ellipsoid} establishes the equivalence between the linear optimization problem and  the separation oracle problem for the same polyhedron. Therefore, the new scheduling problem (\ref{eq:pf:newmaxweight}) can be reduced to the original MaxWeight scheduling problem (\ref{eq:pf:maxweight}) in poly$(L,F)$ time.

We next prove the ``only if'' direction. 

For any input instance $\mathbf{w}\in\mathbb{Z}^D$ in the MaxWeight scheduling problem, construct the input instance $\mathbf{a}\in\mathbb{R}^{DL}$, $\mathbf{b}\in\mathbb{R}^{DL}$ and $\mathbf{c}\in\mathbb{R}^L$ as following.
\begin{align}
&a_{l}^d=(LDB^2+1)w_l,\forall l,d,\notag\\
&b_{l}^d=0,\forall l,d,\quad c_l=1,\forall l.\notag
\end{align}
The above, constant $B$ is the upper bound of the all the link rate $r_l^d$. Suppose that we solve the new scheduling problem in poly$(L,$ $F,\log(\delta^{-1}))$ time under the above input instance. Then we have an $(\mathbf{r}^*,\mathbf{y}^*)$ such that $B((\mathbf{r}^*,\mathbf{y}^*),\delta)\in\mathcal{P}$ and for arbitrary $(\mathbf{r},\mathbf{y})\in\mathcal{P}$,
\begin{align*}
(LDB^2+1)&\sum\limits_{l=1}^{L}w_l y_{l}^* - \sum\limits_{l=1}^{L}\sum\limits_{d=1}^{D}{{r_l^d}^*}^2\geq-\delta+ (LDB^2+1)\sum\limits_{l=1}^{L}w_l y_{l} - \sum\limits_{l=1}^{L}\sum\limits_{d=1}^{D}{r_{l}^d}^2\notag.
\end{align*}
The quantity $y_{l}^*$ and $y_{l}$ derives from $y_l^*=\sum_{d=1}^D {r_l^d}^*$ and $y_l=\sum_{d=1}^D r_l^d$. We prove the following argument by contradiction.
\begin{equation*}
\sum\limits_{l=1}^{L}w_l y_{l}^* \geq \sum\limits_{l=1}^{L}w_l y_{l}, \forall \mathbf{y}\in \Gamma \text{ and } y_l\geq 0, \forall l.
\end{equation*}
Assume that there exists $\mathbf{y}\in \Gamma$ and $y_l\geq 0, \forall l$ such that $\sum\limits_{l=1}^{L}w_l y_{l}^* <\sum\limits_{l=1}^{L}w_l y_{l}$. Then, we have
\begin{align}
&(LDB^2+1)\sum\limits_{l=1}^{L}w_l y_{l}^* <(LDB^2+1)\sum\limits_{l=1}^{L}w_l y_{l}\notag\\
\overset{(a)}{\Rightarrow}&(LDB^2+1)\left[1+\sum\limits_{l=1}^{L}w_l y_{l}^*\right] \leq (LDB^2+1)\sum\limits_{l=1}^{L}w_l y_{l}-\delta\notag\\
\overset{(b)}{\Rightarrow}&(LDB^2+1)\sum\limits_{l=1}^{L}w_l y_{l}^*<\sum\limits_{l=1}^{L}\sum\limits_{d=1}^{D}{{r_l^d}^*}^2-\sum\limits_{l=1}^{L}\sum\limits_{d=1}^{D}{r_l^d}^2-\delta  +(LDB^2+1)\sum\limits_{l=1}^{L}w_l y_{l}\notag\\
\Rightarrow&(LDB^2+1)\sum\limits_{l=1}^{L}w_l y_{l}^* - \sum\limits_{l=1}^{L}\sum\limits_{d=1}^{D}{{r_l^d}^*}^2<  -\sum\limits_{l=1}^{L}\sum\limits_{d=1}^{D}{r_l^d}^2+(LDB^2+1)\sum\limits_{l=1}^{L}w_l y_{l}-\delta\notag,
\end{align}
which is a contradiction. The above, step (a) is based on the assumption that the weight $w_l$, feasible link rate $y_l$, $y_l^*$ are the integers, and that $\delta$ is sufficiently small, step (b) utilizes the definition that $r_l^d\leq B, \forall l,d$. Utilizing the fact that the optimal point of linear optimization lies in the vertex set of the feasible region, the $y_l^*$ is also the optimal solution of the following optimization problem.
\begin{equation*}
\max_{\mathbf{r}}\mathbf{w}^T\mathbf{r},\text{ s.t. } \mathbf{r} \in \mathcal{C}, r_l\geq 0,\forall l.
\end{equation*}
which is clearly the solution of the MaxWeight scheduling problem (\ref{eq:pf:maxweight}). Therefore, the theorem follows.

\end{document}